\documentclass[12pt,aps,tightenlines]{revtex4-2}
\usepackage{amsmath}
\usepackage{amssymb}
\usepackage{bm}
\usepackage[mathscr]{eucal}
\usepackage{graphicx}
\usepackage{psfrag}
\usepackage{subfigure}
\usepackage{color}
\usepackage{wasysym}
\include{graphix}
\usepackage{framed}
\usepackage{enumitem}%
\usepackage{lipsum}
\usepackage{float}
\usepackage{stmaryrd}
\usepackage{MnSymbol}
\definecolor{orange}{rgb}{1,0.5,0}
\usepackage[normalem]{ulem}
\usepackage{cancel}
\usepackage{longtable}

\usepackage{tikz}

\usetikzlibrary{patterns}
\usetikzlibrary{shapes}
\def\normaltwo{\x,{0.7+1/exp(((\x-6)^2)/2)}}

\graphicspath{{figs/}}

\newenvironment{remark}[1][Remark:]{\begin{trivlist}
\item[\hskip \labelsep {\bfseries #1}]}{\end{trivlist}}

\newcommand{\myqed}{\nobreak \ifvmode \relax \else
      \ifdim\lastskip<1.5em \hskip-\lastskip
      \hskip1.5em plus0em minus0.5em \fi \nobreak
      \vrule height0.75em width0.5em depth0.25em\fi}

\newcommand{\mathd}{\mathrm{d}}
\newcommand{\mathe}{\mathrm{e}}

\newcommand{\myRe}{\mathrm{Re}}
\newcommand{\myWe}{\mathrm{We}}
\newcommand{\surften}{\gamma}

\newcommand{\vecf}{\bm{f}}
\newcommand{\vecu}{\bm{u}}

\newcommand{\thetaadv}{\vartheta_a}

\newcommand{\hbl}{h_{bl}}

\newcommand{\rr}{\mathcal{R}_{max}}
\newcommand{\betamax}{\beta_{max}}

\newcommand{\ns}{\lambda_{\mathrm{s}}}

\begin{document}

\title{Bounds on the Spreading Radius in Droplet Impact: The Two-Dimensional Case}
\author{Lennon \'O N\'araigh}
\email{Corresponding author.  Email: onaraigh@maths.ucd.ie}
\affiliation{School of Mathematics and Statistics, University College Dublin, Belfield, Dublin 4, Ireland}
\author{Nicola Young}
\affiliation{School of Mathematics and Statistics, University College Dublin, Belfield, Dublin 4, Ireland}

\date{\today}

\begin{abstract}
We consider the problem of a cylindrical (quasi-two-dimensional) droplet impacting on a hard surface.  Cylindrical droplet impact can be engineered in the laboratory, and a theoretical model of the system can also be used to shed light on various complex experiments involving the impact of liquid sheets.  We formulate a rim-lamella model for the droplet-impact problem.  Using Gronwall's Inequality, we establish theoretical bounds for the maximum spreading radius $\rr$ in droplet impact, specifically $k_1 \myRe^{1/3}-k_2(1-\cos\thetaadv)^{1/2}(\myRe/\myWe)^{1/2}\leq \rr/R_0\leq k_1\myRe^{1/3}$, where $\myRe$ and $\myWe$ are the Reynolds and Weber number based on the droplet's pre-impact velocity and radius $R_0$, $\thetaadv$ is the advancing contact angle (assumed constant in our simplified analysis), and $k_1$ and $k_2$ are constants.  We perform several campaigns of simulations using the Volume of Fluid Method to model the droplet impact, and we find that the simulation results are consistent with the theoretical bounds.
\end{abstract}

\maketitle

\section{Introduction}
\label{sec:intro}

Understanding the physics of droplet on impact on a solid wall is relevant many practical applications, e.g. inkjet printing~\citep{yarin2006drop}, cooling~\citep{yarin2006drop,valluri2015}, and crop spraying~\citep{yarin2006drop,Moghtadernejad2020}.   The phenomenon has been studied using  experimental~\citep{chandra1991collision,antonini2012drop,Riboux2014}, theoretical~\citep{roisman2002normal,gordillo2019theory}, 
and computational~\citep{fukai1998maximum,gunjal2005dynamics,eggers2010drop} methods.
Different impact regimes occur depending on the droplet's Weber number ($\myWe$) and Reynolds number ($\myRe$).  For consistency with previous work  (but going back to Reference~\cite{eggers2010drop}), we use the definitions
\begin{equation}
\myRe=\frac{\rho U_0 R_0}{\mu},\qquad \myWe=\frac{\rho U_0^2 R_0}{\surften},
\end{equation}
where $\rho$ is the liquid density,  $\mu$ is the liquid dynamic viscosity respectively, and $\surften$ is the surface tension.  Also, $U_0$ is the droplet's speed prior to impact and $R_0$ is the droplet radius prior to impact.

In this context, there is a splash parameter $K=\myWe\sqrt{\myRe}$, which determines a threshold above which
splash occurs~\citep{mundo1995droplet,josserand2016drop}.  The threshold value is not universal~\citep{marengo2011drop}, and different experiments have produced different values, a summary of which is provided in Reference~\cite{moreira2010advances}.   Nevertheless, the threshold is of the order of $K=10^3-10^4$.
Just below this threshold, and typically for $\myWe \geq 10^2$ and $\myRe \geq 10^3$~\citep{de2010thickness}, there is `rim-lamella' regime,
in which the droplet flattens and spreads into an axisymmetric structure involving a lamella, with a thicker rim forming at the extremity. A key parameter which characterizes the droplet impact in this regime is the maximum spreading radius, 
denoted here by $\rr$, and is governed by  $\myWe$ and $\myRe$.   

Certainly, droplet spreading is a three-dimensional phenomenon, exhibiting axisymmetry below the splash threshold.   However, this paper focuses instead on cylindrical (quasi-two-dimensional) droplet impacts.   Cylindrical droplets are unstable to the Rayleigh--Plateau instability, and cannot exist naturally over timescales greater than a characteristic breakup time $t_{breakup}\sim [\rho R_0^3/\sigma]^{1/2}$ ($<4\,\mathrm{ms}$ for a $1\,\mathrm{mm}$-radius cylinder of water in air)~\citep{chandrasekhar2013hydrodynamic}.  However, cylindrical droplet impacts can be engineered in the laboratory.  
In the context of droplet impact at high speeds, cylindrical droplets have been generated by confining a liquid bridge between two parallel plates, thereby forming a shape that closely resembles a cylindrical droplet. Impact is achieved by projecting a third plate between the two spaced plates~\citep{field1985studies}. This configuration enables precise observation of the internal dynamics (e.g., via Schlieren imaging), while avoiding  the refraction problems inherent in studies of spherical droplets~\citep{field2012cavitation}. 
Although such studies involve extremely high speeds  ($>100\,\mathrm{m}\cdot\mathrm{s}^{-1}$) where compressibility effects are important, they do show how   cylindrical droplet impacts are realisable in the laboratory and   are therefore a worthwhile subject for theoretical modelling.  Further experiments on incompressible flows have been conducted which  mimic the impact of a cylindrical droplet on a hard surface~\citep{lejeune2018edge,neel2020fines}.   


These experiments motivate the present theoretical study of  idealized two-dimensional droplets.  Another motivation is the computational paper~\cite{tang2024fragmentation}, in which the authors simulate the head-on collision of two liquid cylinders (equivalent to the impact of a  liquid cylinder on a free-slip surface with contact angle $90^\circ$).  These simulations are performed in a full three-dimensional geometry, and fingering is observed post collision, corresponding  to an impact scenario above an effective splash threshold.  Many further  computational studies of two-dimensional droplet impact exist, for instance References~\cite{ding2007diffuse,shin2009simulation,gupta2011two,wu2017dynamics,wu2021decoupled,rafi2022two}.  The present  theoretical analysis provides insights into these computational studies.


In Reference~\cite{naraigh2023analysis}, a two-dimensional energy-budget analysis is proposed.  This allows one to estimate the maximum spreading radius, written in dimensionless terms as $\betamax=\rr/R_0$.  By balancing the energy before impact and at maximum spreading, the following correlation is obtained:
\begin{multline}
\underbrace{1}_{\scriptsize\shortstack{\text{Pre-Impact} \\ \text{Kinetic Energy}}}+\underbrace{\frac{4}{\myWe}}_{\scriptsize\shortstack{\text{Pre-Impact} \\ \text{Surface Energy}}}   =
\underbrace{\frac{2}{\pi\myWe}\left[2\betamax\left(1-\cos\vartheta\right)+ \frac{\pi}{\betamax}\right]}_{\scriptsize\shortstack{\text{Surface Energy} \\ \text{at Maximum Spreading}}}
\\
+\underbrace{\tfrac{2}{\pi}\frac{2a}{\sqrt{\myRe}}\betamax\sqrt{\betamax-1}}_{\scriptsize\shortstack{\text{Viscous Dissipation}\\ \text{in the Boundary Layer}}}+\underbrace{b}_{\scriptsize\text{Head Loss}}.
\label{eq:correlation}
\end{multline}
 Hence, Equation~\eqref{eq:correlation} is an energy balance made dimensionless on the kinetic energy density $(1/2)\rho U_0^2$.  The `head loss' term in Equation~\eqref{eq:correlation} is a key term which reflects energy dissipation due to internal flows which develop in the rim-lamella  structure and which are otherwise not included in a simple energy balance.  The head loss can be modelled as a simple fraction of the initial kinetic energy~\citep{wildeman2016spreading}.  A theoretical justification for this is given 
in Reference~\cite{villermaux2011drop}.  In this way, Equation~\eqref{eq:correlation} contains two free parameters, $a$ and $b$, which have been fitted to the data emanating from numerical simulations~\citep{naraigh2023analysis}.

Equation~\eqref{eq:correlation} has some important asymptotic limits:

\vspace{0.1in}

\begin{itemize}
\item\textit{Inviscid Limit:} For  $\myRe\rightarrow\infty$, Equation~\eqref{eq:correlation} reduces to
\begin{equation}
\frac{\pi}{2}(1-b)\myWe+2\pi=\left[2\beta_{max}(1-\cos\vartheta)+\frac{\pi}{\beta_{max}}\right],
\label{eq:inviscid}
\end{equation}
with exact solution
\begin{equation}
\betamax=\frac{\omega+\sqrt{\omega^2-8\pi(1-\cos\vartheta)}}{4(1-\cos\vartheta)},\qquad \omega=\frac{\pi}{2}(1-b)\myWe+2\pi.
\end{equation}
For $\myWe$ large but finite, this further reduces to:
\begin{equation}
\beta_{max}\approx \frac{\myWe\,\pi(1-b)}{4(1-\cos\vartheta)}.
\label{eq:inviscidWe}
\end{equation}
The equivalent scaling behavior for 3D axisymmetric droplets is:
\begin{equation}
\beta_{max}\approx \sqrt{\frac{4}{1-\cos\vartheta}\left[\tfrac{1}{12}(1-b)\myWe+1\right]},
\end{equation}
hence, $\beta_{max}\sim \myWe$ for 2D Cartesian droplets and $\beta_{max}\sim \myWe^{1/2}$ for 3D axisymmetric droplets.

\item\textit{Finite viscosity, large Weber number:} For $\myWe\rightarrow\infty$ Equation~\eqref{eq:correlation} reduces to:
\begin{equation}
\frac{\pi}{2}(1-b)\approx \frac{2a}{\sqrt{\myRe}}\beta_{max}\sqrt{\beta_{max}-1},
\end{equation}
For $\myRe$ large but finite, this gives $\beta_{max}\sim \myRe^{1/3}$ for 2D Cartesian droplets.  The corresponding result for 3D axisymmetric droplets is $\beta_{max}\sim \myRe^{1/5}$.
\end{itemize}
%
%


\vspace{0.1in}

In the three-dimensional axisymmetric case, a rim-lamella model provides an alternative method for estimating the maximum spreading radius for 3D axisymmetric droplets.  The rim-lamella model is pertinent when the droplet spreads to form a rim-lamella structure (specifically, $\myWe \geq 10^2$, $\myRe \geq 10^3$, and below the splash threshold $K=\myWe\sqrt{\myRe}\apprle 10^3-10^4$).
 In a previous set of papers~\cite{amirfazli2024bounds,naraigh2025bounds}, one of the present authors (with co-authors) was able to place rigorous bounds on the maximum spreading radius for 3D axisymmetric droplets.  The bounds were derived by first formulating a rim-lamella model  and then analysing the resulting set of ordinary differential equations using differential inequalities, in particular, Gronwall's Inequality~\citep{doering1995applied}.   The bounds are found to be consistent with the correlation given in Reference~{roisman2009inertia}, stated here for a Reynolds number and a Weber number based on droplet radius (not diameter):
\begin{equation}
\betamax=1.00\myRe^{1/5}-0.37\myRe^{2/5}\myWe^{-1/2},\qquad
\text{(3D axisymmetric)}.
\label{eq:betamax3D}
\end{equation}
The correlation~\eqref{eq:betamax3D} has been validated extensively via numerical simulation~\citep{wildeman2016spreading} and experimental measurements.

\subsection{Aim of the paper}

The main aim of the present work is  to establish a result analogous to Equation~\eqref{eq:betamax3D} for two-dimensional droplets.  For this purpose,  we formulate a rim-lamella model for two-dimensional droplets.  Within the framework of the rim-lamella model, we show:
\begin{equation}
k_1 \myRe^{1/3}-k_2 \myRe^{1/2}\myWe^{-1/2}(1-\cos\thetaadv)^{1/2} \leq \betamax\leq k_1 \myRe^{1/3}, \qquad \text{(2D Cartesian)},
\label{eq:bounds2D}
\end{equation}
where $k_1$ and $k_2$ are constants to be determined, and $\thetaadv$ is the advancing contact angle.     
We  verify the bounds~\eqref{eq:bounds2D} using  computational fluid dynamics 
simulations of droplet impact, based on a volume-of-fluid computational model for the multiphase flow.

As a by-product of this analysis, we find that the inviscid rim-lamella model ($\myRe=\infty$) is exactly solvable in two dimensions (unlike in the 3D axisymmetric case).  Therefore, a final aim of the work is to rigorously analyse this special case, and to present special closed-form analytical solutions to the rim-lamella model.

\subsection{Plan of the paper}

In Section~\ref{sec:model} we derive the rim-lamella model from first principles.  The result is a set of coupled ordinary differential equations for the rim area $V$, the greatest extent of the lamella $R$, and the lamella height $h$.  In Section~\ref{sec:inviscid} we examine the inviscid limit of the model, where we derive a closed-form expression for $R(t)$, the variation of $R$ with respect to time.  Crucially, we find a closed-form expression for $R_{max}$, and show that it scales linearly with Weber number, a result consistent with the energy-budget analysis~\eqref{eq:inviscidWe}.  In Section~\ref{sec:viscous} we extend the analysis to the viscous case.  In this case, an exact closed-form solution is not possible; however, we use an analysis of the rim-lamella model based on Gronwall's inequality to establish the bounds~\eqref{eq:bounds2D}.  In Section~\ref{sec:numerics} we validate the bounds using numerical simulations of two-dimensional droplet impact within the framework of  the Volume of Fluid method.  Discussion and concluding remarks are presented in Section~\ref{sec:conc}.

\vspace{0.1in}

\begin{remark}In this paper, we use the symbol $V$ for rim area, to emphasize the similarity of theoretical model with previous work on 3D axisymmetric droplets, where $V$ was used to denote a rim volume.
\end{remark}

\section{Mathematical Model}
\label{sec:model}

We introduce a rim-lamella model for two-dimensional droplets, based on the analogous one for three-dimensional axisymmetric droplets introduced in Reference~\cite{naraigh2025bounds} (but see also References~\cite{eggers2010drop,gordillo2019theory}). 
\begin{figure}[htb]
\centering
\begin{tikzpicture}[scale=1.1, transform shape]
\draw[-,black,line width=0.8mm] (0,0) -- (12,0);
\foreach \x in {-1,...,22}
\draw (0.5+0.5*\x,0) -- (0.5+0.5*\x+0.5*0.707,-0.5*0.707);
\draw[->,black,line width=0.3mm] (6,0) -- (13,0);
\draw (13, -0.2) node[below] {$x$};
\draw[->,black,line width=0.3mm] (6,0) -- (6,3);
\draw (6, 3) node[right] {$z$};
\draw (6 ,0.25) node[right] {$O$};
%
\draw[black,dashed,line width=0.3mm,domain=2:10] plot (\normaltwo) node[right] {};
%
\draw[black,dashed,line width=0.3mm] (1.75,0.7) -- (10.5,0.7);
%
\draw [black,line width=0.3mm,domain=-45:140] plot ({10.8+0.5*cos(\x)}, {0.35+0.5*sin(\x)});
\draw [black,line width=0.3mm,domain=32:228] plot ({1.4+0.5*cos(\x)}, {0.35+0.5*sin(\x)});
%
\draw [<->,red,line width=0.5mm] (2.5,0) -- (2.5,0.7);
\draw (2.5, 0.4) node[right] {${\color{red}{h}}$}; 
%
%
\draw [>-<,red,line width=0.5mm] (3.5,-0.1) -- (3.5,0.3);
\draw (3.7, 0.3) node[right] {${\color{red}{\hbl}}$}; 
%
%
\draw [->,red,line width=0.5mm] (6,-0.7) -- (10.5,-0.7);
\draw (9.5,-0.7) node[below] {${\color{red}{R}}$}; 
\draw [-,red,line width=0.5mm] (10.45,0) -- (10.45,0.7);
\draw [-,red,line width=0.55mm] (1.8,0) -- (1.8,0.7);
%
%
\draw [dotted,red,line width=0.5mm] (10.45,0) -- (10.45,-1.5);
\draw [dotted,red,line width=0.5mm] (11.2,0) -- (11.2,-1.5);
\draw [<->,red,line width=0.5mm] (10.45,-1.3) -- (11.2,-1.3);
\draw (10.85, -1.4) node[below] {${\color{red}{2a}}$}; 
%
\draw (3, 1.5) node[rectangle,draw,below] {Lamella};
\draw (1, 1.5) node[rectangle,draw,below] {Rim};
\draw [black,dashed,line width=0.5mm] (11.2,0) -- (11.7,1);
\draw (11.5, 0.3) node[right] {$\pi-\thetaadv$};
\draw [->,blue,line width=0.5mm] (11,0.8) -- (12.5,0.8);
\draw (13,0.9) node[above] {{\textcolor{blue}{$U=\mathd R/\mathd t$}}};
%
\draw [->,blue,line width=0.5mm] (7.5+1.5,0.45) -- (8.2+1.5,0.45);
\draw [-,blue,line width=0.5mm] (9.73,0.2) -- (9.73,0.73);
\draw [blue,dashed,line width=0.5mm] (9,0) -- (9,0.7);
\draw (9.6,0.8) node[above] {{\textcolor{blue}{$u_o$}}};
%
%
%
%
\draw[color=black,line width=0.3mm] (1.75,0.2) -- (2.4,0.2);
\draw[color=black,line width=0.3mm] (2.6,0.2) -- (3.8,0.2); 
\draw[color=black,line width=0.3mm] (4.3,0.2) -- (6.1,0.2);
\draw[color=black,line width=0.3mm] (6.5,0.2) -- (10.5,0.2);
\draw [->,black,line width=0.5mm] (9,2.5) -- (10,2.5);
\draw (10, 2.5) node[below] {$x$};
\draw [->,black,line width=0.5mm] (9,2.5) -- (9,3.5);
\draw (9, 3.5) node[left] {$z$};
\fill (9,2.5) circle (2pt); 
\end{tikzpicture}
\caption{Schematic diagram showing the cross-section of an axi-symmetric rim-lamella structure.  Dashed curved line:  true lamella height $h(x,t)=(t+t_0)^{-1}f[(x/(t+t_0)]+h_{PI}(t)$, as given by Equation~\eqref{eq:ht1}.    Dashed straight line:  the remote asymptotic approximation $h(x,t)=h_{init}(\tau+t_0)/(t+t_0)+h_{PI}(t)$.}
\label{fig:schematic}
\end{figure}
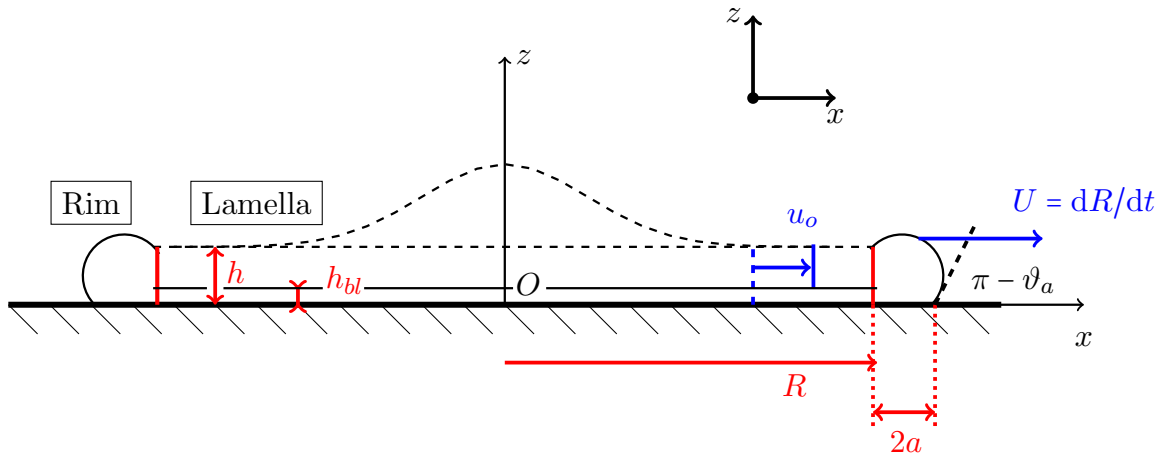
The schematic diagram for rim-lamella droplet spreading scenario is shown in Figure~\ref{fig:schematic}.
We work through the different elements of the model, starting with the flow in the lamella.

\vspace{0.1in}

\begin{remark}
In this work, we will use a rim-lamella model to place bounds on the maximum value of $R$.  However, from Figure~\ref{fig:schematic}, it can be seen that $\rr=R+2a$, where $\rr$ is the quantity of interest.  However, in Section~\ref{sec:viscous} we will demonstrate that $2a$ is negligibly small compared to $R$, and hence can be ignored.
\end{remark}

\subsection{Flow in the Lamella}
 The flow far from the substrate  is denoted by $u_o$ (the outer flow).  Similar to the 3D axisymmetric case, this is assumed to be a hyperbolic flow  and is given (in 2D Cartesian coordinates) by the expression
\begin{equation}
u_o(x,t)=\frac{x}{t+t_0},
\label{eq:uouter}
\end{equation}
where $t_0$ is a model parameter which describes the onset time of the flow, prior to the formation of the rim-lamella structure.   The assumption of this hyberbolic flow profile in the lamella is justified by earlier simulations of two-dimensional droplet impact~\citep{naraigh2023analysis}.

A boundary layer forms close to the substrate, in which the fluid velocity transitions from $u_o$ to zero, across a distance $\hbl$, the boundary-layer thickness.  From boundary-layer 
theory~\citep{white2011fluid}, the boundary layer thickness scales as $\hbl\propto \sqrt{\nu x/u_o}$, where $\nu=\mu/\rho$ is the kinematic viscosity of the liquid.  Given the functional form of $u_o$, the $x$-dependence cancels out, to give $\hbl\propto \sqrt{\nu (t+t_0)}$.  We allow for an additional degree of freedom in the problem by taking $\hbl\propto\sqrt{\nu (t+t_1)}$, where $t_1$ describes the time of formation of the boundary layer.  Following boundary-layer theory~\citep{white2011fluid}, we describe the variation of the flow in the vertical direction using a transition function, such that:
\begin{equation}
u(x,z,t)=u_o(x,t)F(z),
\label{eq:utotal}
\end{equation}
where $F(z=0)=0$ and $F(z)=1$ for $z\geq \hbl$.  For simplicity, and following References~{naraigh2025bounds,eggers2010drop} for the 3D axisymmetric case, we use a step function for the transition function, such that:
\begin{equation}
F(z)=\begin{cases} 0,& z<\hbl,\\
                                                       1,& z>\hbl.\end{cases}
\label{eq:step}
\end{equation}
Finally, we use the incompressibility condition $\partial_x u+\partial_z w=0$ to obtain an expression for $w(x,z,t)$:
%
%
\begin{equation}
w(x,z,t)=\begin{cases} 0,& z<\hbl,\\
                      -\frac{z-\hbl}{t+t_0},&z>\hbl.\end{cases}
\end{equation}

\subsection{Lamella Height}

To model the lamella height $h(x,t)$, we assume the kinematic condition, such that the interface $h(x,t)$ moves with the flow.
Hence, the height of the lamella $h(x,t)$ satisfies:
\begin{equation}
\frac{\partial h}{\partial t}+u\big|_{z=h}\frac{\partial h}{\partial x}=w\big|_{z=h}.
\label{eq:ht}
\end{equation}
In the remainder of this section, we assume we are in \textit{Phase 1} of the motion, such that $h(R,t)>\hbl$, and such that $(u,w)_{z=h}\neq 0$.  We discuss \textit{Phase 2} of the motion, wherein $h(R,t)<\hbl$ and $(u,w)_{z=h}=0$ in Section~\ref{sec:viscous}.
 Thus, Equation~\eqref{eq:ht} becomes:
\begin{equation}
\frac{\partial h}{\partial t}+\frac{x}{t+t_0}\frac{\partial h}{\partial x}=w\big|_{z=h}=-\frac{h-\hbl}{t+t_0}.
\label{eq:ht1}
\end{equation}
Equation~\eqref{eq:ht1} has an exact solution by the method of characteristics:
\begin{equation}
h(x,t)=\frac{\tau+t_0}{t+t_0}f\left(x\frac{\tau+t_0}{t+t_0}\right)+
\underbrace{\tfrac{2}{3}\left[\hbl(t)\frac{t+t_1}{t+t_0}-\hbl(\tau)\frac{\tau+t_1}{t+t_0}\right]}_{=h_{PI}(t)}\qquad t\geq \tau,
\label{eq:ht1_sln}
\end{equation}
where $f$ is set by the initial conditions.  At late times, $f$ is taken to be a constant function, which is the so-called remote asymptotic solution~\citep{roisman2009inertia}.  Here, we use the parameter $\tau$ to denote the initial time $\tau$ at which point the rim-lamella structure seen in Figure~\ref{fig:schematic} forms.

\subsection{Mass Conservation}

We derive an ODE for the total volume of the lamella in what follows.  We omit the redundant dimension  out of the plane of the page, which carries with it a factor of $\lambda$, this being the length of the idealized cylindrical droplet.  Hence, we work with the rim area $V$, the lamella area $2\int_0^R \mathd x\int_0^{h(x,t)}\mathd z$ and the total area $V_{tot}$, such that:
\begin{equation}
V_{tot}=V+2\int_0^R \mathd x\int_0^{h(x,t)}\mathd z,
\label{eq:Vtot}
\end{equation}
We use $\mathd V_{tot}/\mathd t=0$ and Equation~\eqref{eq:ht1} to get:
\begin{equation}
\frac{\mathd V}{\mathd t}=2\left[\frac{R}{t+t_0}\left(1-\frac{\hbl}{h}\right)-\dot R\right]h,
\label{eq:dVdt}
\end{equation}
where $R$ denotes the greatest extent of the lamella at time $t$, and where we henceforth use the simplified notation
\begin{equation}
h\equiv h(t)\equiv h(R,t).
\label{eq:hequiv}
\end{equation}
  By mass conservation, Equation~\eqref{eq:dVdt} has exact solution
\[
V=V_{tot}-2g\left(R\frac{\tau+t_0}{t+t_0}\right)-2\left(R\frac{\tau+t_0}{t+t_0}\right)h_{PI}(t),
\]
where $g(\eta)=\int_0^\eta f(\eta)\mathd \eta$.

\subsection{Momentum}

In a similar way, we derive an ODE for the total momentum of the lamella.  We work with one half of the rim momentum $P$, and one half of the lamella momentum 
$\rho\int_0^R \mathd x\int_0^{h(x,t)}\mathd z$, such that the total momentum in (say) the right-hand half of the rim-lamella structure in Figure~\ref{fig:schematic} is:
%
%
%
\begin{equation}
P_{tot}=P+\rho\lambda\int_0^R \mathd x\int_0^h u \mathd z,
\end{equation}
where again, $\lambda$ is the length of the idealized cylindrical droplet in the plane of the page.
When $h>\hbl$ (Phase 1) this becomes:
\begin{equation}
P_{tot}=P+\underbrace{\rho\lambda\int_0^R \frac{x}{t+t_0}\left[h(x,t)-\hbl(t)\right]\mathd x}_{=P_{lam}}.
\end{equation}
By direct differentiation, we have:
\begin{equation}
\frac{\mathd P_{lam}}{\mathd t}
=\rho\lambda\left[\frac{R}{t+t_0}\left[h(R,t)-\hbl(t)\right]\left(\dot R-\frac{R}{t+t_0}\right)-\tfrac{1}{4}\frac{\hbl R^2}{(t+t_0)(t+t_1)}\right].
\label{eq:newt1}
\end{equation}
By Newton's Second Law, we have:
\begin{equation}
\frac{\mathd P_{tot}}{\mathd t}=F_{ext}=-\lambda\,\surften(1-\cos\thetaadv). 
\label{eq:newt2}
\end{equation}
Here, $\thetaadv$ is the advancing contact angle, assumed for the present purposes to be constant.
Hence, combining Equation~\eqref{eq:newt1} and~\eqref{eq:newt2} we have:
\begin{equation}
-\lambda\surften(1-\cos\thetaadv)-\frac{R}{t+t_0}\rho\lambda\left[h(R,t)-\hbl(t)\right]\left(\dot R-\frac{R}{t+t_0}\right)+\tfrac{1}{4}\frac{\rho\lambda\hbl R^2}{(t+t_0)(t+t_1)}=\frac{\mathd P}{\mathd t}.
\end{equation}
We use $\mathd P/\mathd t=\rho\lambda U (\mathd V_{1/2}/\mathd t)+\rho\lambda V_{1/2}(\mathd U/\mathd t)$ ($V_{1/2}$ is half the rim area) and $U\equiv \dot R$ to re-write the momentum equation in a final form:
\begin{equation}
\rho V\frac{\mathd U}{\mathd t}=2\rho\bigg\{\left[h-\hbl(t)\right]\left(u_0-U\right)^2+U^2\hbl\bigg\}+\tfrac{1}{2}\frac{\rho \hbl R^2}{(t+t_0)(t+t_1)}-2\surften(1-\cos\thetaadv),
\label{eq:momentum}
\end{equation}
where $u_0=R/(t+t_0)$, and where we have divided out by the redundant lengthscale $\lambda$.

\subsection{Summary of the model equations}

We gather up the model equations here in one place.  We emphasize that the equations are for Phase 1 of the motion, wherein $h(t)>\hbl(t)$:
\begin{subequations}
\begin{eqnarray}
\frac{\mathd V}{\mathd t}&=&2\left[ u_0\left(h-\hbl\right)-U h\right],\\
\frac{\mathd R}{\mathd t}&=&U,\\
V\frac{\mathd U}{\mathd t}&=&2\left[(u_0-U)^2\left(h-\hbl\right)+U^2\hbl\right]\nonumber\\
&\phantom{=}&\phantom{aaaaa}+\tfrac{1}{2}\frac{\hbl R^2}{(t+t_0)(t+t_1)}-2\frac{\gamma}{\rho}(1-\cos\thetaadv),
\end{eqnarray}%
\label{eq:rl_basic}%
\end{subequations}%
where $h$ is given by Equation~\eqref{eq:ht1_sln} and~\eqref{eq:hequiv}, and $u_0=R/(t+t_0)$.  We analyse Equation~\eqref{eq:rl_basic} in the following sections.

\section{Inviscid Case}
\label{sec:inviscid}

In this section we look at the rin-lamella model~\eqref{eq:rl_basic} in the inviscid case, which can be obtained by shrinking the boundary layer $\hbl$ to zero.  Hence, the model simplifies:
\begin{subequations}
\begin{eqnarray}
\frac{\mathd V}{\mathd t}&=&2\left(u_0-U\right)h,\\
\frac{\mathd R}{\mathd t}&=&U,\\
V\frac{\mathd U}{\mathd t}&=&2\left(u_0-U\right)^2h-2\gamma\left(1-\cos\thetaadv\right).
\end{eqnarray}
\label{eq:rl_basic}%
\end{subequations}
Unlike the corresponding inviscid rim-lamella model in the 3D axisymmetric case~\citep{amirfazli2024bounds}, Equation~\eqref{eq:rl_basic} has an exact solution, which we present here.

\subsection{Exact solution}
As in the 3D axisymmetric model, we introduce the velocity defect $\Delta=u_0-U$.  Equation~\eqref{eq:rl_basic} can then be recast in terms of an ODE for $\Delta$:
\begin{equation}
\frac{\mathd \Delta}{\mathd t}+\frac{\Delta}{t+t_0}=-\frac{2h}{V}\left(\Delta^2-c^2\right),
\label{eq:delta1}
\end{equation}
where 
\begin{equation}
c= \sqrt{ \frac{\surften}{\rho h}(1-\cos\thetaadv)} 
\end{equation}
is the Taylor--Culick speed ($c$ is time-dependent, through $h$).  We further recast Equation~\eqref{eq:delta1} as an ODE for $V\Delta$:
\begin{equation}
\frac{\mathd }{\mathd t}(V\Delta)+\frac{V\Delta}{t+t_0}=2hc^2=2(\gamma/\rho)(1-\cos\thetaadv).
\end{equation}
Thus, there is an exact solution for $V\Delta$:
\begin{equation}
V\Delta=V_{init}\Delta_{init}\left(\frac{\tau+t_0}{t+t_0}\right)+\frac{\gamma}{\rho}\left(1-\cos\thetaadv\right)\left[t+t_0-\frac{(\tau+t_0)^2}{t+t_0}\right].
\end{equation}
We recall the definition of $\Delta$, to obtain:
\begin{equation}
\frac{\mathd R}{\mathd t}-\frac{R}{t+t_0}=-\frac{V_{init}}{V}\Delta_{init}\left(\frac{\tau+t_0}{t+t_0}\right)-\frac{\gamma}{\rho}\frac{\left(1-\cos\thetaadv\right)}{V}\left[t+t_0-\frac{(\tau+t_0)^2}{t+t_0}\right].
\end{equation}
This can be further re-written as:
\begin{equation}
\frac{\mathd }{\mathd t}\left(\frac{R}{t+t_0}\right)=-\frac{V_{init}}{V}\frac{\Delta_{init}}{t+t_0}\left(\frac{\tau+t_0}{t+t_0}\right)-\frac{\gamma}{\rho}\frac{\left(1-\cos\thetaadv\right)}{V}\left[1-\left(\frac{\tau+t_0}{t+t_0}\right)^2\right].
\end{equation}
To make further progress, we use the remote asymptotic solution for $h(R,t)\equiv h$:
\begin{equation}
h(R,t)=\left(\frac{\tau+t_0}{t+t_0}\right)h_{init},
\end{equation}
hence $V=V_{tot}-2R[(\tau+t_0)/(t+t_0)]h_{init}$.  We also group together the terms $R[(\tau+t_0)/(t+t_0)]$ as $\eta$.  
Thus, we have:
\begin{equation}
\frac{\mathd\eta}{\mathd t}=-\Delta_{init}\frac{V_{init}}{V_{tot}-2\eta h_{init}}\left(\frac{\tau+t_0}{t+t_0}\right)^2-\frac{\gamma(\tau+t_0)}{\rho}\frac{\left(1-\cos\thetaadv\right)}{V_{tot}-2\eta h_{init}}\left[1-\left(\frac{\tau+t_0}{t+t_0}\right)^2\right].
\label{eq:separable}
\end{equation}
This is a separable ODE.
Following the steps outlined in Appendix~\ref{sec:app:mathanal_invisc}, Equation~\eqref{eq:separable} can be solved to yield:
\begin{equation}
Y=X- \sqrt{\epsilon^2 X^2+A \left(X^3+X-2X^2\right)+B \epsilon\left(X^2-X\right)},\qquad X\geq 1,
\label{eq:exact1}
\end{equation}
where
\begin{subequations}
\begin{eqnarray}
X&=& \frac{t+t_0}{\tau+t_0},\\
Y&=&\frac{R}{V_{tot}/(2h_{init})},\\
A&=&\frac{c_*^2(\tau+t_0)^2}{V_{tot}},\\
c_*^2&=&4(h_{init}/V_{tot})(\surften/\rho)(1-\cos\thetaadv),\\
B&=&\frac{4\Delta_{init}(\tau+t_0)h_{init}}{V_{tot}}\\
\epsilon&=&\frac{V_{init}}{V_{tot}}.
\end{eqnarray}%
\label{eq:AB}%
\end{subequations}%
Equation~\eqref{eq:exact1} is the required exact solution of the rim-lamella model.

\subsection{Maximum Spreading Radius}

The maximum spreading occurs at $\mathd R/\mathd t=0$, hence $\mathd Y/\mathd X=0$.  Based on Equation~\eqref{eq:exact1}, this occurs when:
\begin{equation}
4\left[\epsilon^2 X^2+A \left(X^3+X-2X^2\right)+ \epsilon B\left(X^2-X\right)\right]=
\left[3A X^2+2X\left(\epsilon^2+\epsilon B-2A\right)+\left(A-\epsilon B\right)\right]^2.
\end{equation}
Assuming the maximum spreading occurs at  large value of $t$ (hence, $X$), this gives $X\approx 4/(9A)$,
a result which will be verified \textit{a posteriori}.  Correspondingly,
\begin{equation}
Y_{max}\approx (4/9)A^{-1}-\sqrt{A (4/9)^3 A^{-3}}=\tfrac{4}{27}A^{-1}.
\end{equation}
By using the explicit definition of $A$ from Equation~\eqref{eq:AB} we have (details in Appendix~\ref{sec:app:mathanal_invisc}):
\begin{equation}
\frac{R_{max}}{R_0}\approx \tfrac{1}{27} \frac{V_{tot}^2}{ h_{init} U_0^2 (\tau+t_0)^2 R_0 }\left(\frac{V_{tot}}{2h_{init}R_0}\right)\frac{\myWe}{1-\cos\thetaadv}.
\label{eq:rmaxWe}
\end{equation}
In order for the  large-$X$ approximation to be valid, we require $A$ to be small, and hence, $\myWe$ to be large.  Thus, Equation~\eqref{eq:rmaxWe} is valid in the large Weber-number limit.  This scaling behaviour ($\rr/R_0\sim \myWe$, for $\myWe\gg 1$) is exactly that is seen in previous published works on two-dimensional droplet impact, and is consistent with the energy-budget analysis~\eqref{eq:inviscidWe}.

\subsection{Parameter Estimation}

In order to make a systematic comparison between Equation~\eqref{eq:rmaxWe} and both the correlation~\eqref{eq:inviscidWe} and numerical simulations, it is necessary to estimate the pre-factors in Equation~\eqref{eq:rmaxWe}.

Following Reference~{amirfazli2024bounds} for the rim-lamella model in 3D axisymmetry (inviscid case), we estimate $h_{init}/(\tau+t_0)$ as $U_0$, hence:
\begin{equation}
\frac{R_{max}}{R_0}\sim \tfrac{1}{27} \frac{V_{tot}^2}{ h_{init}^3 R_0 }\left(\frac{V_{tot}}{2h_{init}R_0}\right)\frac{\myWe}{1-\cos\thetaadv},\qquad \myWe\gg 1.
\label{eq:Rmax1}
\end{equation}
We estimate $h_{init}$ by a technique also used in Reference~{amirfazli2024bounds} for the rim-lamella model in 3D axisymmetry.  Hence, we assume that -- just prior to the generation of the rim -- the droplet flattens into a sheet (effectively, a rectangle in two dimensions, of sides of length $2R_{init}$ and $h_{init}$).  The corresponding rim area is $V_{init}=0$.  An energy balance at this instant gives:
\begin{equation}
\text{Kinetic Energy of sheet}+\text{Surface Energy of sheet}=\gamma(2\pi R_0)+\tfrac{1}{2}\rho V_{tot}U_0^2.
\label{eq:energyest}
\end{equation}
In more detail, this expression is:
\begin{multline}
2\times \tfrac{1}{2}\rho\int_0^R\int_0^{h_{init}}\left[\left(\frac{x}{\tau+t_0}\right)^2+\left(\frac{z}{\tau+t_0}\right)^2 \right]\mathd x\,\mathd z+\left[2R_{init}(1-\cos\thetaadv)+2h_{init}\right]\gamma\\=
\gamma(2\pi R_0)+\tfrac{1}{2}\rho V_{tot}U_0^2.
\end{multline}
We evaluate the integral, and use the fact that $V_{tot}=2R_{init}h_{init}=\pi R_0^2$, hence $R_{init}=V_{tot}/(2h_{init})$.   We further divide both sides by $\rho U_0^2 R_0^2$.  This gives:
\begin{multline}
\tfrac{1}{3} \frac{1}{R_0^2 U_0^2(\tau+t_0)^2}\left[ \left(\frac{V_{tot}}{2h_{init}}\right)^3 h_{init} +\left(\frac{V_{tot}}{2h_{init}}\right)h_{init}^3\right]\\
+\frac{1}{\myWe}\left[ 2\left(\frac{V_{tot}}{2h_{init}}\right)\left(1-\cos\thetaadv\right)+2h_{init}\right]=\frac{2\pi}{\myWe}+\tfrac{1}{2}\pi
\end{multline}
We multiply both sides by $h_{init}^2$ and go over to the large-$\myWe$ limit.  This gives;
\begin{equation}
\tfrac{1}{3} \underbrace{\frac{h_{init}^2}{U_0^2(\tau+t_0)^2}}_{=1}\frac{1}{R_0^2}\left[ \left(\frac{V_{tot}}{2h_{init}}\right)^3 h_{init} +\left(\frac{V_{tot}}{2h_{init}}\right)h_{init}^3\right]\sim \tfrac{1}{2}	\pi h_{init}^2,\qquad \myWe\gg 1.
\end{equation}
Hence:
\begin{equation}
\left(\frac{V_{tot}}{2h_{init}}\right)^2\frac{\cancel{V_{tot}}}{2\cancel{h_{init}}} \cancel{h_{init}}+\frac{\cancel{V_{tot}}}{2\cancel{h_{init}}}\cancel{h_{init}}h_{init}^2\sim \tfrac{3}{2}\cancel{(\pi R_0^2)}h_{init}^2,\qquad \myWe\gg 1,
\end{equation}
which simplifies to:
\begin{equation}
\frac{V_{tot}^2}{8}\sim  h_{init}^4,\qquad \myWe\gg 1.
\end{equation}
We go back up to Equation~\eqref{eq:Rmax1}, with $h_{init}^4\sim V_{tot}^2/8$ for large $\myWe$, hence:
\begin{equation}
\frac{R_{max}}{R_0}\sim \tfrac{4\pi}{27} \frac{\myWe}{1-\cos\thetaadv},\qquad \myWe \gg 1.
\label{eq:rmaxRL}
\end{equation}
To two significant figures, this gives $R_{max}/R_0\sim 0.47\,\myWe/(1-\cos\thetaadv)$, valid for $\myWe \gg 1$.

\subsection{Comparisons}

We compare the result~\eqref{eq:rmaxRL} with other results from the literature.  First, we look at the energy budget~\eqref{eq:correlation} in the limit of large $\myWe$ and $\myRe=\infty$, which gives:
\begin{equation}
\frac{R_{max}}{R_0}\sim \tfrac{\pi(1-b)}{4}  \frac{\myWe}{1-\cos\thetaadv}.
\label{eq:correlationWe}
\end{equation}
Following the energy-budget results in Reference~\cite{wildeman2016spreading} (albeit for three-dimensional axisymmetric droplets), we take $b=1/2$ as per the `head loss' argument therein.  To two significant figures, this gives $R_{max}/R_0\sim 0.39 \myWe/(1-\cos\thetaadv)$.
In contrast, if we view $b$ as a fitting parameter and fit the correlation~\eqref{eq:correlationWe} to the DNS data in Reference~{wu2021decoupled}, we get $b=0.536$ (the fitting of the DNS data to the correlation~\eqref{eq:correlationWe} is done in Reference~\cite{naraigh2023analysis}), hence 
$R_{max}/R_0\sim 0.36 \myWe/(1-\cos\thetaadv)$.

We take $b=1/2$ as the reference case, due to the simplicity of the head-loss argument and its theoretical basis~\citep{villermaux2011drop}.  Thus, there is a relative error of $21\%$ between the prefactor in the rim-lamella model and the reference case.  The discrepancy can be traced back to the estimate~\eqref{eq:energyest} for the initial kinetic energy: in order for the prefactor in the rim-lamella model and the reference-case energy budget to agree, the kinetic energy estimate on the LHS of Equation~\eqref{eq:energyest} would need to be increased by a factor of $1.12$.  Thus, the discrepancy in the rim-lamella model arises due to a (slight) under-estimation of the initial kinetic energy (12\%) in the LHS of Equation~\eqref{eq:energyest}.

\section{Viscous case}
\label{sec:viscous}

Having completely characterized the inviscid case down to deriving an exact solution for the spreading radius $R(t)$, we return to the viscous case, recalled here as Equation~\eqref{eq:rl_basic}.  These equations are valid for Phase 1 of the motion, in which the height of the boundary layer is still small compared to the film height $h$.  The aim of this section is to analyse Phase 1 of the motion, and also to study Phase 2, in which the height of the boundary layer exceeds $h$.  In this second phase, the form of the Equations~\eqref{eq:rl_basic} drastically simplify.  

\subsection{Recasting of Equations for the Rim-Lamella System}

We recall Equation~\eqref{eq:rl_basic}(c) for the momentum:
\begin{equation}
V\frac{\mathd U}{\mathd t}=2\left[(u_0-U)^2\left(h-\hbl\right)+U^2\hbl\right]+\tfrac{1}{2}\frac{\hbl R^2}{(t+t_0)(t+t_1)}-2\frac{\gamma}{\rho}(1-\cos\thetaadv).
\label{eq:RLvisc1}
\end{equation}
We re-write:
\begin{equation}
(u_0-U)^2(h-\hbl)+U^2\hbl = \left[ u_0\left(1-\frac{\hbl}{h}\right)-U\right]^2h + u_0^2 (\hbl/h)\left(1-\frac{\hbl}{h}\right).
\end{equation}
We introduce $\overline{u}=u_0[1-(\hbl/h)]$.
Thus, Equation~\eqref{eq:RLvisc1} becomes:
\begin{equation}
V\frac{\mathd U}{\mathd t}=2\left(\overline{u}-U\right)^2h+\underbrace{u_0^2 (\hbl/h)\left(1-\frac{\hbl}{h}\right)+\tfrac{1}{2}\frac{\hbl R^2}{(t+t_0)(t+t_1)}}-2\frac{\gamma}{\rho}(1-\cos\thetaadv).
\label{eq:underbrace}
\end{equation}
In a previous work on the analogous 3D axisymmetric rim-lamella model~\citep{naraigh2025bounds}, a term analogous to the one with the underbrace in Equation~\eqref{eq:underbrace} was omitted, on the basis that it was negligibly small (as justified in Reference~{gordillo2019theory}) and also, led to a mathematically tractable set of equations.  We follow the same approach here, and we therefore work with a modified momentum equation:
\begin{equation}
V\frac{\mathd U}{\mathd t}=2\left[ \left(\overline{u}-U\right)^2 -c^2\right]h,
\label{eq:RLvisc2}
\end{equation}
where $c^2=[\gamma/(\rho h)](1-\cos\thetaadv)$, and where $c$ is the Taylor--Culick speed.  Therefore, for the purposes of this section, we examine the following rim-lamella model:
\begin{subequations}
\begin{eqnarray}
\frac{\mathd V}{\mathd t}&=&2\left(\overline{u}-U\right)h,\\
\frac{\mathd R}{\mathd t}&=&U,\\
V\frac{\mathd U}{\mathd t}&=&2\left[ \left(\overline{u}-U\right)^2 -c^2\right]h.
\end{eqnarray}%
\label{eq:rl_reg}%
\end{subequations}%

\subsection{Initial Conditions}

Initial conditions apply at the onset of Phase 1.  The model has built-in an implied set of initial conditions that give rise to rim generation.
 At time $t=R_0/U_0=\tau$, we take the rim volume to be zero, as in the analogous work on 3D axisymmetric rim-lamella models~\citep{amirfazli2024bounds,naraigh2025bounds}.
This gives a natural way to parametrize the rim generation phenomenon, since, by taking $V(\tau)=0$, we obtain (from Equation~\eqref{eq:rl_reg}(c)):
\begin{equation}
\left(\overline{u}-U\right)^2h - \frac{\surften\left(1-\cos\vartheta_{ap}\right)}{\rho} =0.
\label{eq:Utau1}
\end{equation}
Hence:
\begin{equation}
U(t=\tau)=\overline{u}-\sqrt{ \frac{\surften\left(1-\cos\vartheta_{ap}\right)}{\rho h}}, 
\label{eq:Utau2}
\end{equation}
or
\begin{equation}
U(t=\tau)=\frac{R_{init}}{\tau+t_0}\left(1-\frac{\hbl}{h_{init}}\right)-c(\tau).
\label{eq:sqrt2}
\end{equation}

The values $R_{init}$ and $h_{init}$ describe the initial lamella, i.e. just prior to the formation of the rim.  For the inviscid case (Section~\ref{sec:inviscid}),  it is important to describe carefully the dependence of these parameters on $\myWe$, as this has implications for the value of the prefactor in $R_{max}/R_0\sim \myWe$ for $\myWe \gg 1$.
For the viscous case, this seems less important.   For instance, in Reference~\cite{roisman2002normal}  $R_{init}$ and $h_{init}$ are obtained using an energy-budget analysis, whereas in Reference~\cite{eggers2010drop} a geometric argument is used.  In these studies, the final dependence of the spreading radius on $\myRe$ and $\myWe$ is insensitive to the approach used to set the initial conditions.    We use a similar geometric argument here: we assume that the droplet assumes a `pancake' shape at time $\tau$, with initial height $h_{init}=R_0/2$.  This gives $R_{init}=V_{tot}/(2h_{init})$, where $V_{tot}=\pi R_0^2$ is the total (conserved) volume.

\subsection{Phase 2 of the motion}

In Phase 2 of the motion, the boundary layer exceeds the height of the lamella.  In the present simplified mathematical model, all flow inside the lamella then stops and Equation~\eqref{eq:ht} for the interface height reduces to $\partial h/\partial t=0$, hence $h=\text{Const.}$.  We use $t_*$ to denote the onset time of Phase 2; $t_*$ is obtained from Equation~\eqref{eq:ht1_sln} by solving:
\begin{equation}
h(t_*)=\hbl(t_*).
\end{equation}
Here, we work with the remote asymptotic solution, such that $f(\cdot)=\text{Const.}$ in Equation~\eqref{eq:ht1_sln}.
In this case, Equations~\eqref{eq:rl_reg} simplify greatly:
\begin{subequations}
\begin{eqnarray}
\frac{\mathd V}{\mathd t}&=& -2U h_*,\\
\frac{\mathd R}{\mathd t}&=& U,\\
V\frac{\mathd U}{\mathd t}&=&2\left[U^2h_* -\frac{\gamma}{\rho}(1-\cos\thetaadv)\right].
\end{eqnarray}%
\label{eq:RLvisc_P2}%
\end{subequations}%
Equations~\eqref{eq:RLvisc_P2} have the simple fixed-point solution
\begin{equation}
U=-c_*,\qquad V=2Uh_*,
\end{equation}
which remains valid until $R=R(t_*)-c_* t$ touches down to zero (here, $R(t_*)$ is the value of $R(t)$ at the onset of Phase 2). Also,
\begin{equation}
c_*=\sqrt{ \frac{\gamma}{\rho h_*}(1-\cos\thetaadv)}
\end{equation}
is the Taylor--Culick retraction speed.  We note in passing the use of $c_*$ for a different speed in Section~\ref{sec:inviscid}, and rely on the reader's judgment to distinguish these, based on the context.

\subsection{Analysis of the Equations of Motion using inequalities}

Unlike in the inviscid case the viscous rim-lamella model~\eqref{eq:rl_reg} does not admit an exact solution.  However, we can characterize the solution using differential inequalities, similar to what was done in in the 3D axisymmetric case in Reference~\cite{naraigh2025bounds}.  We summarize the approach here, and refer the interested reader to Appendix~\ref{sec:app:mathanal_visc} for the lengthy calculations which underpin this summary.

\begin{equation}
\Delta=u_0\left(1-\frac{\hbl}{h}\right)-U.
\end{equation}
Hence,  Equations~\eqref{eq:RLvisc1} become:
\begin{subequations}
\begin{eqnarray}
\frac{\mathd V}{\mathd t}&=&2\Delta h,\\
\frac{\mathd R}{\mathd t}&=&U,\\
V\frac{\mathd U}{\mathd t}&=&2(\Delta^2-c^2)h.
\end{eqnarray}%
\label{eq:RLvisc2}%
\end{subequations}%
By direct computation, we get:
\begin{equation}
\frac{\mathd\Delta}{\mathd t}+\frac{\Delta}{t+t_0}\left(1-\frac{\hbl}{h}\right)=-2(\Delta^2-c^2)(h/V)-\frac{u_0}{t+t_0}\frac{\hbl}{h}\underbrace{\left[2\left(1-\frac{\hbl}{h}\right)+\tfrac{1}{2}\frac{t+t_0}{t+t_1}\right]}_{=\Phi(t)\geq 0}.
\end{equation}
We assume $V(t=\tau)=0$.  We use Gronwall's Inequality to deduce (Appendix~\ref{sec:app:mathanal_visc}):
\begin{equation}
\Delta \leq (\gamma/\rho)(1-\cos\thetaadv)(h/V)I_h(t).
\label{eq:Deltavisc1}
\end{equation}
where
\begin{equation}
I_h(t)=\int_\tau^t \frac{\mathd t}{h}.
\end{equation}
We use $\Delta=[R/(t+t_0)][1-(\hbl/h)]-(\mathd R/\mathd t)$ to re-write Equation~\eqref{eq:Deltavisc1} as:
\begin{equation}
\frac{\mathd R}{\mathd t}-\frac{R}{t+t_0}\left(1-\frac{\hbl}{h}\right)\geq -2(\gamma/\rho)(1-\cos\thetaadv)(h/V)I_h(t).
\label{eq:Deltavisc2}
\end{equation}
This can be re-written as:
\begin{equation}
\frac{\mathd }{\mathd t}(Rh)\geq -2(\gamma/\rho)(1-\cos\thetaadv)(h^2/V)I_h(t).
\label{eq:Deltavisc3}
\end{equation}
We note that $V=V_{tot}-2Rh$ and  identify $\eta=Rh$.  Hence, we re-write Equation~\eqref{eq:Deltavisc3} as:
\begin{equation}
\frac{\mathd \eta}{\mathd t}\geq -2(\gamma/\rho)(1-\cos\thetaadv)\frac{h^2 I_h(t)}{V_{tot}-2\eta}.
\end{equation}
This can be solved to give:
\begin{equation}
V_{tot}\eta-\eta^2\geq V_{tot}\eta_{init}-\eta_{init}^2 -(\gamma/\rho)(1-\cos\thetaadv)\underbrace{\left[2\int_\tau^t h^2 I_h(t)\mathd t\right]}_{=\Delta G(t)}.
\label{eq:Deltavisc4}
\end{equation}
Equation~\eqref{eq:Deltavisc4} is a quadratic inequality.  Critical points occur at $\eta_{\mathrm{cr},\pm}(t)$, where:
\begin{equation}
\eta_{\mathrm{cr},\pm}(t)=\tfrac{1}{2}V_{tot}\pm \left[(\gamma/\rho)(1-\cos\thetaadv)\right]^{1/2}[\Delta G(t)]^{1/2}
\label{eq:etacr}
\end{equation}
(we write  $\eta_{\mathrm{cr},\pm}(t)$ to indicate that the critical points are time-dependent).
Since $\eta(t)=R(t)h(t)$, we have $R(t)h(t)\leq \eta_{\mathrm{cr},+}(t)$.
Similarly, with the negative sign chosen, we get $R(t)h(t)\geq \eta_{\mathrm{cr},-}(t)$.  Since this is true for all $t$, we must have $R(t_*)h(t_*)\geq \eta_{\mathrm{cr},-}(t_*)$, hence
\begin{equation}
R(t_*)\geq \frac{\eta_{\mathrm{cr}-}(t_*)}{h(t_*)}\stackrel{\text{Eq.~\eqref{eq:etacr}}}{=} \tfrac{1}{2}\frac{V_{tot}}{h(t_*)}-\underbrace{\left[(\gamma/\rho)(1-\cos\thetaadv)\right]^{1/2}\frac{[\Delta G(t_*)]^{1/2}}{h(t_*)}}.
\label{eq:Rineq1}
\end{equation}
We apply the following observations to inequality~\eqref{eq:Rineq1}:
\begin{itemize}[noitemsep]
\item Since the term with the underbrace is proportional to $\gamma^{1/2}$, and $\gamma\propto \myWe^{-1}$, the RHS of the inequality is positive, for sufficiently large $\myWe$;
\item By definition, we have $\max_t R(t)\geq R(t_*)$.
\end{itemize}
Putting these observations together, we have:
\begin{equation}
\max_t R(t)\geq R(t_*)\geq \tfrac{1}{2}\frac{V_{tot}}{h_*}-\left[(\gamma/\rho)(1-\cos\thetaadv)\right]^{1/2}\frac{[\Delta G(t_*)]^{1/2}}{h_*}>0.
\label{eq:Rineq2}
\end{equation}
Here, the last inequality is true for $\myWe$ sufficiently large; without this, the string of inequalities would be vacuous.

For an upper bound, we reuse the argument in Reference~\cite{naraigh2025bounds}.  We have:
\begin{equation}
2\underbrace{R_{max}}_{=R(t_{max})}h(t_{max})=V_{tot}-V(t_{max})\leq V_{tot}.
\end{equation}
Hence,
\begin{equation}
R_{max}\leq \frac{V_{tot}}{2 h(t_{max})}.
\end{equation}
If the maximum occurs in Phase 1, then $h(t)$ is monotone-decreasing, hence $h(t_{max})\geq h_*$, hence
$1/h(t_{max})\leq 1/h_*$, and
\begin{equation}
R_{max}\leq \frac{V_{tot}}{2 h_*}.
\label{eq:Rstarmax}
\end{equation}
If the  maximum occurs in Phase 2, then $h(t)$ is a  constant function and equal to $h_*$, so it is still the case that $R_{max}\leq V_{tot}/(2h_*)$.  Hence, in both cases, inequality~\eqref{eq:Rstarmax} is true. Hence, by a sandwich result, we have:
\begin{equation}
\frac{\tfrac{1}{2}V_{tot} - \left[(\gamma/\rho)(1-\cos\thetaadv)\right]^{1/2}[\Delta G(t_*)]^{1/2}}{h_*} \leq \max_t R(t)\leq \frac{V_{tot}}{2 h_*}.
\label{eq:Rineq3}
\end{equation}

\subsection{Evaluation of the bounds}

To make further progress, we study $h_*$ in more detail, to extract its $\myRe$-dependence.  For this purpose, we solve Equation~\eqref{eq:ht1} numerically.  We assume the remote asymptotic approximation, such that $h(x,t)$ no longer depends on space, and such that Equation~\eqref{eq:ht1} becomes an ODE, specifically:
\begin{equation}
\frac{\mathd h}{\mathd t}=-\frac{h-\hbl}{t+t_0},\qquad \tau<t<t_*,\qquad \hbl=\alpha \sqrt{\nu (t+t_1)}.
\label{eq:ht2}
\end{equation}
As per the previous discussion in this section, we use the initial condition $h(\tau)=R_0/2$.  Equation~\eqref{eq:ht2} can be solved explicitly; however it is just as straightforward to solve it numerically in Matlab.  We use the stiff solver \texttt{ode15s}, which readily handles the transition which occurs when $\mathd h/\mathd t$ jumps abruptly to zero at $t=t_*$.  We furthermore solve Equation~\eqref{eq:ht2} using dimensionless variables, such that all lengthscales are normalized by $R_0$, and all velocity scales by $U_0$.  The fluid density $\rho$ sets the scale for mass.  Equation~\eqref{eq:ht2} can then be nondimensionalized by formally replacing $\nu$ by $\myRe^{-1}$, and by implicitly taking $h\rightarrow h/R_0$, $t\rightarrow t U_0/R_0$, and similarly for $\tau$, $t_0$, and $t_1$.  The dimensionless parameter $\alpha$ is unchanged in this process.

\begin{figure}
	\centering
		\subfigure[]{\includegraphics[width=0.45\textwidth]{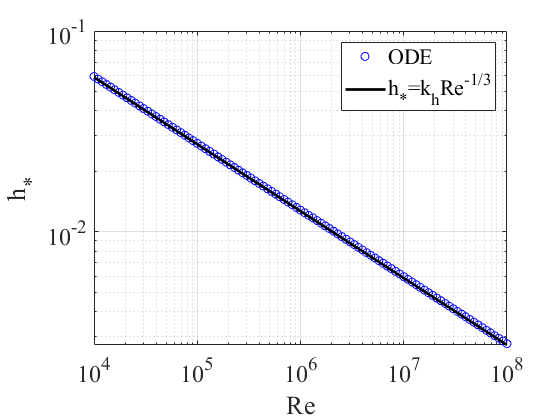}}
		\subfigure[]{\includegraphics[width=0.45\textwidth]{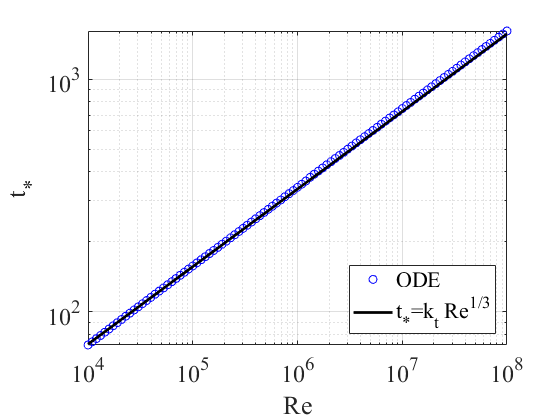}}
		\subfigure[]{\includegraphics[width=0.45\textwidth]{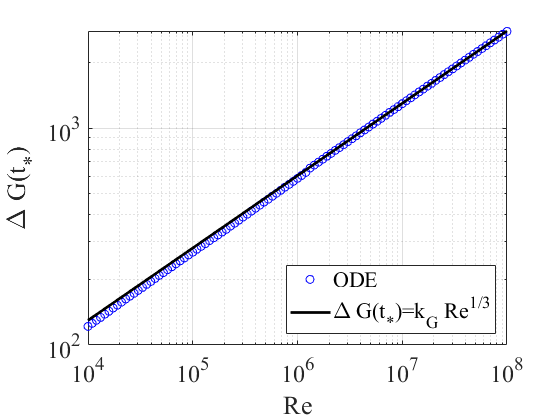}}
		\caption{Numerical analysis of Equation~\eqref{eq:ht2} showing the asymptotic dependence of $h_*$, $t_*$, and $\Delta G(t_*)$ on $\myRe$, for large values of $\myRe$.}
	\label{fig:h_star}
\end{figure}

We assume that $\tau=U_0/R_0$ marks the onset of the rim-lamella phase (or, $\tau=1$ in dimensionless variables).  Hence, Equation~\eqref{eq:ht2} contains three free parameters: $\alpha$, $t_0$, and $t_1$.  We choose these parameters via non-linear optimization, with the objective function to be described in what follows.  For the time being, we report these values here:
\begin{equation}
\alpha=0.6908,\qquad t_0=2.0,\qquad t_1 =2.8907.
\label{eq:abc}
\end{equation}
%
%
%
%
Results are shown in Figure~\ref{fig:h_star}.
From the figure, we find:
\begin{equation}
h_*=k_h \myRe^{-1/3},\qquad t_*=k_t \myRe^{1/3},\qquad \myRe \gg 1,
\label{eq:hstar1}
\end{equation}
and
\begin{equation}
\Delta G =k_G \myRe^{1/3},\qquad \myRe \gg 1,
\label{eq:hstar2}
\end{equation}
where $k_h$, $k_t$, and $k_G$ are given in Table~\ref{tab:kvals}.
\begin{table}
	\centering
		\begin{tabular}{|c|c|c|}
		\hline
		$k_h$ & $k_t$ & $k_G$ \\
		\hline 
			 1.2668 & 3.3680 &  6.0312\\
		\hline
		\end{tabular}
		\caption{Numerical values for the parameters $k_h$, $k_t$, and $k_G$}
		\label{tab:kvals}
\end{table}
We apply the results~\eqref{eq:hstar1}--\eqref{eq:hstar2} to the bound~\eqref{eq:Rineq3}.  This gives the sandwich result
\begin{equation}
k_1\myRe^{1/3}-k_2 (1-\cos\thetaadv)^{1/2} (\myRe/\myWe)^{1/2} \leq \frac{\max R(t)}{R_0}\leq k_1\myRe^{1/3} ,\qquad \myRe \gg 1,
\label{eq:correlation_final}
\end{equation}
where $k_1=\pi /(2 k_h)$, and $k_2=(2 k_G)^{1/2}/k_h$.

\subsection{Correction due to the rim}

The results so far have involved expressions for $R_{max}$, the maximum extent of the lamella.
However, what is of interest is $\rr$, being the maximum spreading radius of the rim-lamella
structure. The maximum spreading radius can be written as $\rr = R_{max} + 2a$, where $2a$ is the
footprint of the rim, and where a can be estimated from Figure~\ref{fig:schematic} using a circular-segment construction as:
\begin{equation}
a=\sin\thetaadv\sqrt{\frac{2V}{2\thetaadv-\sin(\thetaadv)}}
\end{equation}
where $V$ is the area of the rim (for an in-depth presentation of this construction, see Reference~\cite{amirfazli2024bounds}).  Since $V\leq V_{tot}$, the correction to $\rr$ does not exceed a term trivially proportional to $\myRe^0$
%
%
%
and hence, can be ignored in the remaining calculations. For the avoidance of doubt, in the remainder of this article we use $R_{max}$ as the quantity of interest.

\section{Numerical Simulations}
\label{sec:numerics}

\begin{figure}[ht]
    \centering
    \begin{tikzpicture}
				 \node[anchor=south west, inner sep=0] (image) at (0,0) {\includegraphics[width=12cm]{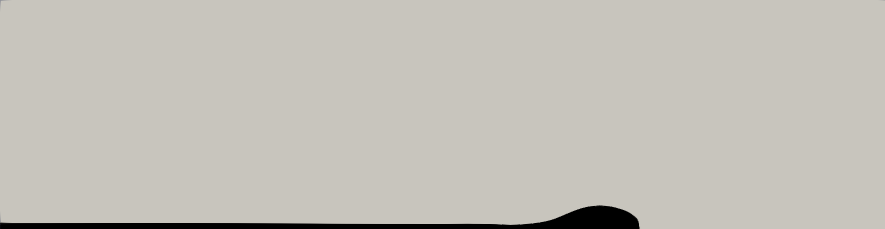}};
				\begin{scope}[shift={(0,0)}]
            \draw[->, thick, blue] (0,0) -- (12.5,0) node[below] {$x$};  
            \draw[->, thick, blue] (0,0) -- (0,3.5) node[left] {$z$}; 
						%
						\node at (6,-0.5) {Slip BC on $\vecu$, contact-line BC on $\alpha$};
						\node at (6,3.5) {No slip};
						\node[rotate=90] at (-0.5,1.5) {Symmetry plane};
            \node[rotate=90] at (12.5,1.5) {Zero gradient};
						%
						%
						%
	          \foreach \x in {0,0.5,...,12} {
            \draw[thick,blue] (\x,0) -- ++(0.25,-0.25);
						}
						%
						%
            \fill[red!50] (0,0.7) arc[start angle=-90,end angle=90,radius=0.7];
						%
            \node[draw, fill=white] at (2.7,1.5) {\shortstack{$t=0:$\\$\alpha=1$ \\ $\vecu = (0,0,-1)$ m/s}};
	          \draw[->] (1.1,1.5) -- (0.7,1.5);
            \node[draw, fill=white] at (10,2) {$\alpha = 0$};
						%
						%
						\draw[thick, green, dashed] (8.7,0) -- (8.7,1); 
	          \draw[thick,green] (8,0) arc[start angle=180,end angle=90,radius=0.7];
						\node[draw, fill=white] at (9.2,0.5) {$\thetaadv$};
						\node[draw, fill=white] at (11.7,0.5) {$L_x$};
            \node[draw, fill=white] at (0.5,3.2) {$L_z$};
        \end{scope}
    \end{tikzpicture}
    \caption{Schematic diagram showing the computational setup.  The `constantContactAngle' boundary condition is applied to the scalar field $\alpha$ which sets the contact angle (construction shown in green in the figure) to the value $\thetaadv$ at each timestep.  The figure shows $\thetaadv=\pi/2$.}
		\label{fig:schematic_cfd}
\end{figure}
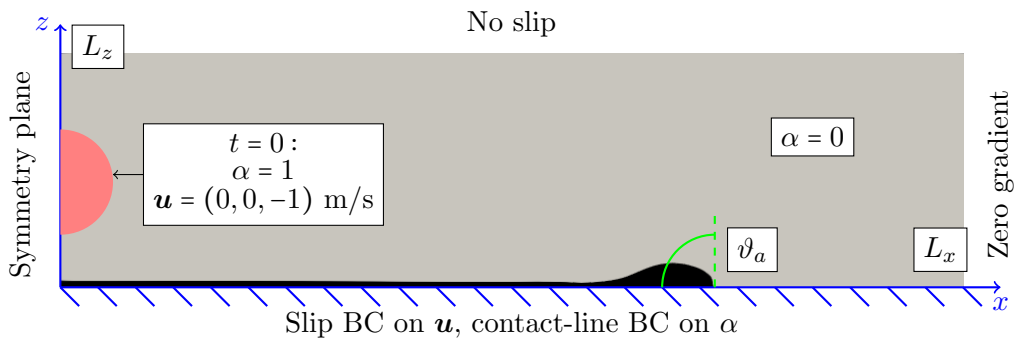

To validate the correlation~\eqref{eq:correlation_final}, we simulate the impact of a droplet on a hard surface in a two-dimensional Cartesian geometry, as shown in Figure~\ref{fig:schematic_cfd}.  We use the Volume of Fluid (VOF) method to simulate the gas-liquid two-phase flow.  To implement the VOF method, we use the OpenFOAM computational fluid dynamics software.

In this context, a scalar field  $\alpha$ (the volume fraction) is used to represent the relative volume of the liquid phase in each computational cell.  Thus,  $\alpha = 1$ indicates a cell filled entirely with liquid, and $\alpha = 0$ represents a cell filled entirely with gas.  Intermediate values denote a mixed region, and the level set $\alpha=1/2$ represents the interface.   The volume fraction is advected using a modified form of the continuity equation, ensuring it remains bounded, with $\alpha\in [0,1]$. The governing equation for $\alpha$ is therefore:
\begin{subequations}
\begin{equation}
\frac{\partial \alpha}{\partial t} + \nabla \cdot (\alpha \vecu) + \nabla \cdot [\alpha(1 - \alpha)\vecu_r] = 0.
\end{equation}
Here, $\vecu$ is the velocity field, and $\vecu_r$ is an artificial compression velocity used to sharpen the interface~\citep{cifani2016comparison}.

The time-evolution of the velocity field $\vecu$ is given by a one-fluid formulation of the Navier-Stokes equations, obtained by using the effective properties (density $\rho$ and viscosity $\mu$) calculated as weighted averages based on $\alpha$. The momentum equation takes the form:
\begin{equation}
\rho\left(\frac{\partial\vecu}{\partial t} + \vecu\cdot\nabla\vecu \right)= -\nabla p + \nabla \cdot \left[\mu\left( \nabla \vecu+\nabla\vecu^T\right)\right] +  \vecf_\surften,
\label{eq:OFmom}
\end{equation}
where $\vecf_\surften$ denotes the surface-tension force, modelled here as a bulk forcing term (as opposed to a surface-only forcing term), with the model given by Brackbill's Continuum Surface Force~\citep{brackbill1992continuum}.  As we are isolating the effect of surface tension on the rim-lamella dynamics, the gravity term is excluded from Equation~\eqref{eq:OFmom}.   Finally, the velocity field is incompressible:
\begin{equation}
\nabla\cdot\vecu=0.
\label{eq:OFincomp}
\end{equation}
\label{eq:OFall}
\end{subequations}

\subsection{Numerical Implementation}

We solve Equations~\eqref{eq:OFall} numerically using a rectangular domain of size $L_x\times L_z$ as shown in Figure~\ref{fig:schematic_cfd}.  We take $L_x=0.046\,\mathrm{m}$, $L_z=0.012\,\mathrm{m}$ and the initial droplet size $R_0=0.003\mathrm{m}$.  To simulate droplet impact, the droplet is seeded with an initial velocity $\vecu=(0,0,-1)\,\mathrm{m}\cdot\mathrm{s}^{-1}$.
We discretize the domain
using a simple block mesh in OpenFOAM with $2250$ gridpoints in the $x$-direction and $750$ gridpoints in the $z$-direction.   A mesh-refinement study is described below.
No grading of the mesh is applied in the $x$-direction.  However, simple grading is applied in the $z$-direction such that the grid spacing $\Delta z$ of the cell closest to the bottom wall ($z=0$) is half that of the cell at the top wall.  This is to better capture the flow inside the lamella at late times.   To capture the droplet at maximum spreading, some simulations require a larger value of $L_x$.  In these cases, $L_x$ is increased while keeping the resolution $\Delta x=L_x/n_x$ the same.

We use the PIMPLE algorithm to solve for the pressure and implement the incompressibility condition~\eqref{eq:OFincomp}.  In this context, the Navier--Stokes equations are solved explicitly, with 1 outer corrector and 3 inner correctors per timestep.  Full details of the OpenFOAM case have been posted in the accompanying GitHub repository~\citep{droplet2D}.

Boundary and initial conditions are described in Figure~\ref{fig:schematic_cfd}.  We use the `constantContactAngle' boundary condition on the scalar field $\alpha$ to enforce an equilibrium contact angle at the triple point.  We also use a Navier slip boundary condition on the velocity field at the bottom wall.  A standard no-slip condition is applied at the top wall.  We describe the slip boundary condition in more detail below.   
The third dimension into the plane of the page is chosen to be very small so as to make the simulation quasi-two-dimensional.  Boundary conditions of type `empty' are therefore applied to the corresponding faces.  

We solve Equations~\eqref{eq:OFall} with standard transport properties for an air-water system, as summarized in Table~\ref{tab:aw}.
\begin{table}
	\centering
		\begin{tabular}{|c|c|}
		\hline
	  $\nu$ (air) & $ 1.5\times 10^{-5}\,\mathrm{m}^2\cdot\mathrm{s}^{-1}$ \\
		$\nu$ (water) & $ 1.0\times 10^{-6}\,\mathrm{m}^2\cdot\mathrm{s}^{-1}$ \\
		$\rho$ (air) & $ 1.225\,\mathrm{kg}\cdot\mathrm{m}^{-3}$ \\
		$\rho$ (water) & $ 1000\,\mathrm{kg}\cdot\mathrm{m}^{-3}$ \\
		$\surften$  & $ 0.072\,\mathrm{N}\cdot\mathrm{m}^{-1}$\\
		\hline
		\end{tabular}
		\caption{Transport properties used in the model}
		\label{tab:aw}
\end{table}
Based on these properties, we have $\myWe=41.67$ and $\myRe=3,000$.  In order to study the effect of the Weber number and the Reynolds number on the maximum spreading radius, we will vary $\surften$ and $\nu$ (water) systematically in what follows.

\subsection{Slip Boundary Condition}

In a two-dimensional setting, and using the coordinate system in Figure~\ref{fig:schematic_cfd}, the Navier slip boundary at the bottom wall is expressed by:
\begin{equation}
u_x+\ns\frac{\partial u_x}{\partial z}=0,\qquad z=0,
\label{eq:ns1}
\end{equation}
where $u_x$ denotes the velocity component in the $x$-direction and $\ns$ is the slip length.  If we denote $u_0$ as the tangential velocity ($=u_x$) on the lower boundary and $u_1$ as the tangential velocity at the adjacent cell center, then the discretized version of Equation~\eqref{eq:ns1} reads:
\begin{equation}
u_0+\ns\frac{\left(u_0-u_1\right)}{d/2}=0,
\end{equation}
where $d$ is the size of the cell adjacent to the lower wall.  Re-arranging gives:
\[
u_0=\left(1-f\right)u_1,\qquad f=\frac{d}{d+2\lambda},
\]
The parameter $f$ is inputted into OpenFOAM and describes the slip ($f$ corresponds to `valueFraction' in the  boundary condtion of type `partialSlip').   A value $f=1$ corresponds to no slip and a value $f=0$ corresponds to perfect slip.

\begin{figure}
	\centering
		\includegraphics[width=0.7\textwidth]{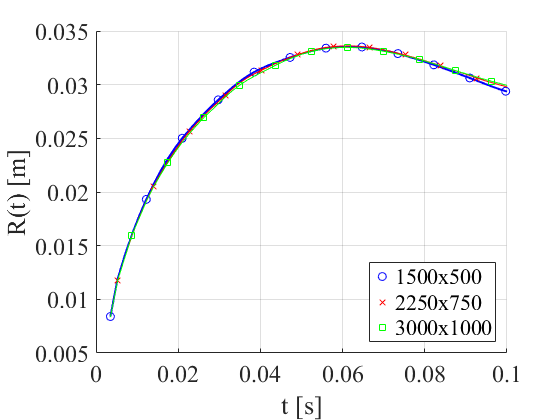}
		\caption{Mesh refinement study at constant slip length, $\ns=8.32\times 10^{-6}\,\mathrm{m}$, showing the triple point $R(t)$ as a function of time.  Slip length: $\ns=8.32\times 10^{-6}\,\mathrm{m}$.  Other parameters as in Tables~\ref{tab:aw}--\ref{tab:factor}.  }
	\label{fig:convergence_slip}
\end{figure}

As noted in Reference~\cite{legendre2015comparison}, the VOF method applied to moving contact lines exhibits poor convergence unless a slip condition with constant slip length is used.  For this reason, we use a slip length $\ns=8.32\times 10^{-6}\,\mathrm{m}$ (hence, $(\ns/R_0)\times 100<0.28\%$)
We show a mesh-refinement study at constant $\lambda$ (hence, varying $f$, see Table~\ref{tab:factor}) in Figure~\ref{fig:convergence_slip}.
\begin{table}
	\centering
		\begin{tabular}{|c|c|c|c|}
		\hline
		         & Resolution & $\Delta z$                       & $f$ \\
		\hline 
		Study 1  &   $1500\times 500$         &$1.66\times 10^{-5}\,\mathrm{m}$ & 0.5 \\
		Study 2  &   $2250\times 750$         &$1.11\times 10^{-5}\,\mathrm{m}$ & 0.4 \\
		Study 3  &   $3000\times 1000$        &$5.45\times 10^{-6}\,\mathrm{m}$ & 0.333 \\
		\hline
		\end{tabular}
		\caption{Mesh refinement study at constant slip length, $\ns=8.32\times 10^{-6}\,\mathrm{m}$}
		\label{tab:factor}
\end{table}
This figure shows that the mesh of size $2250\times 750$ cells is sufficient to produce a converged simulation, and hence accurately to describe the spreading behavior.  This amounts to a large number of cells for a two-dimensional simulation (over 1.6 million), this large number of cells is needed to accurately resolve the thin lamella structure which forms at late times.  This can be seen e.g. in Figure~\ref{fig:schematic_cfd}, which overlays annotation on a snapshot of the simulation, to produce a schematic diagram.  Since the simulations are two-dimensional, the present approach of achieving converged simulations simply by adding more cells throughout the computational domain is valid.  For equivalent three-dimensional simulations, a more computationally efficient approach would be required, e.g. local grid refinement near the bottom wall, or adaptive mesh refinement.

\vspace{0.1in}

\begin{remark}
We emphasize that a constant contact-angle model is used throughout the simulations, which rather unphysically forces the contact angle to its equilibrium value even when the triple-point velocity is non-zero.  However, this simplified approach is sufficient to validate the bounds~\eqref{eq:correlation_final}, and hence, to achieve one of the main aims of the paper.  To maintain continuity with the other parts of the paper, we use the symbole $\thetaadv$ to denote this constant contact angle.
\end{remark}

\subsection{Results}

\begin{figure}
	\centering
		\includegraphics[width=0.7\textwidth]{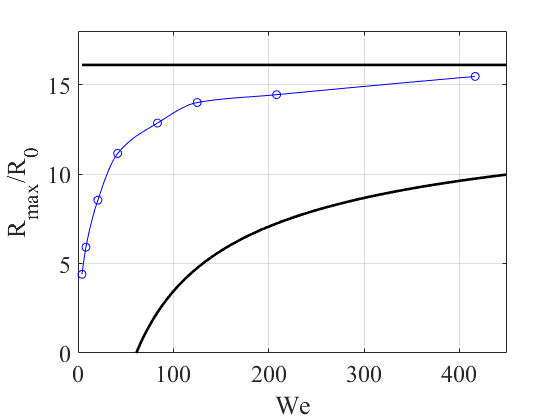}
		\caption{Blue Curve: Dependence of $\betamax=\rr/R_0$ on Weber number, at fixed Reynolds number and contact angle $\thetaadv=90^\circ$ (CAMPN1).  Apart from the varying Weber number (through the parameter $\surften$), the parameters are as given in Table~\ref{tab:aw}. Black Solid Lines: The bounds~\eqref{eq:correlation_final}.}
	\label{fig:results0}
\end{figure}

\begin{figure}
	\centering
		\includegraphics[width=0.7\textwidth]{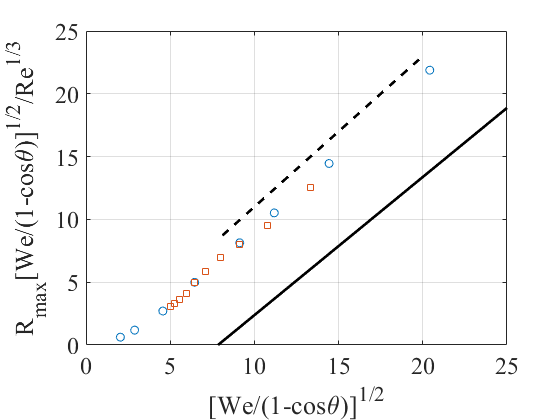}
		\caption{Collapse of results on to a single curve.  Circles: Varying $\myWe$ at fixed $\thetaadv$ (CAMPN1)  Squares: Varying $\thetaadv$ at fixed $\myWe$ (CAMPN2).  All other parameters as in Table~\ref{tab:aw}.  Black solid line: the lower bound given by~\eqref{eq:correlation_final}.  Broken solid line: Trend line, giving the slope $Y=k_1X$ and $k_1=1.24$.}
	\label{fig:collapse}
\end{figure}

In a first campaign of simulations (CAMPN1), we vary $\sigma$ (hence $\myWe$) while keeping $\myRe$ and $\thetaadv=90^\circ$ fixed, and we show the results in Figure~\ref{fig:results0}.  
In a second campaign of simulations (CAMPN2), we fix $\myWe$ and $\myRe$ as per Table~\ref{tab:aw} and vary $\thetaadv$ through a range $\thetaadv\in [40^\circ,130^\circ]$.  The results from this campaign, together with CAMPN1 at varying Weber number all collapse onto a single curve (Figure~\ref{fig:collapse}).  For large values of $X=[\myWe/(1-\cos\thetaadv)]^{1/2}$, the variable $Y=\betamax X/\myRe^{1/3}$ depends linearly on $X$, as evidenced by the trend line in the figure.  Given the putative correlation $\betamax=k_1 \myRe^{1/3}$ valid at sufficiently high Weber number, we identify the linear relationship between $X$ and $Y$ as $Y=k_1 X + \text{Const.}$ (the constant may be $\myRe$-dependent), and we fit the value $k_1=1.24$.

 With this value in mind, we revisit the model~\eqref{eq:ht2} and perform non-linear least-squares fitting over the parameters $(\alpha,t_0,t_1)$ such that $\pi/(2k_h)=k_1$ is forced to take the value $1.24$.  This then provides us with the given parameter values for $\alpha$, $t_0$, and $t_1$ in Equation~\eqref{eq:abc}.  With these parameter values, we confirm that the DNS results fall comfortably inside the bounds~\eqref{eq:correlation_final} implied by the rim-lamella model.

\begin{figure}
	\centering
		\includegraphics[width=0.7\textwidth]{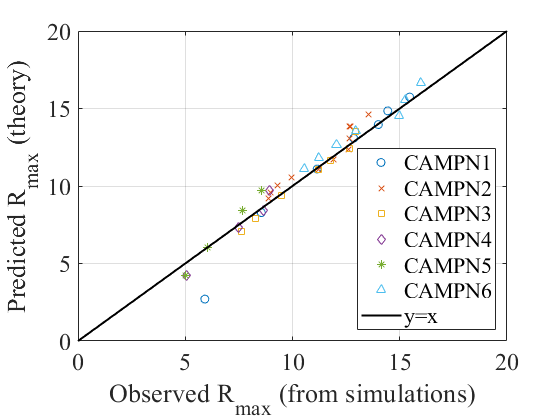}
		\caption{Observed values of $\rr$ (horizontal axis) compared to predicted values of 
		$\rr$ using the correlation~\eqref{eq:predicted}.}
	\label{fig:universal_curve}
\end{figure}

In another campaign of simulations (CAMPN3), we  vary $\nu$ (hence $\myRe$) while keeping $\myWe$ and $\thetaadv=90^\circ$ fixed.  We are able to fit the resulting values of $\betamax$ to a curve of the form
\begin{equation}
\betamax=k_1 \myRe^{1/3}-k_0 (1-\cos\thetaadv)^{1/2}(\myRe/\myWe)^{1/2}.
\label{eq:predicted}
\end{equation}
The value of $k_1$ is fixed from CAMPN1-CAMPN2 as $k_1=1.24$; we therefore view $k_0$ as a single fitting parameter, which is estimated based solely on the results from CAMPN3 ($k_0=0.80$).  Then, to visualize the results, we plot the values of $\betamax$ from CAMPN3 on the horizontal axis, and the predicted values of $\betamax$ (from Equation~\eqref{eq:predicted}) on the vertical axis.  Crucially, we superimpose on this plot the actual versus predicted values from CAMPN1-CAMPN2, and the result is a near-collapse of the data onto a straight line of slope $45^\circ$.  Results from further simulation campaigns (CAMPN4-CAMPN6) are also plotted on the same graph (Figure~\ref{fig:universal_curve}), and a full synopsis of the data emanating from all campaigns is given in Tables~\ref{tab:camp1}--\ref{tab:camp2}.  
Notably, Campaign 5 uses the Diffuse Interface Method for the droplet impact simulations (DIM, see Reference~\cite{naraigh2023analysis} for the method); the results are consistent with the other campaigns which use the VOF method in OpenFOAM.
The overall result is the validation of the universal scaling behavior~\eqref{eq:correlation}, in the rim-lamella regime of droplet spreading.


\begin{table}
	\centering
\begin{tabular}{|p{2cm}|p{1.5cm}|p{1.5cm}|p{2.5cm}|p{1.5cm}|p{2cm}|}
\hline
Campaign & $\myWe$ & $\myRe$ & Method & $\thetaadv$ & $\rr/R_0$ \\
\hline 
\hline
			1 &	416.7	&3000	&OpenFOAM 	&$90^\circ	$&15.46\\
			1	&208.3	&3000	&OpenFOAM 	&$90^\circ	$&14.44\\
			1	&125.0	&3000	&OpenFOAM 	&$90^\circ	$&14.00\\
			1	&83.33	&3000	&OpenFOAM 	&$90^\circ	$&12.85\\
			1	&41.67	&3000	&OpenFOAM 	&$90^\circ	$&11.15\\
			1	&20.83	&3000	&OpenFOAM 	&$90^\circ	$&8.546\\
			1	&8.333	&3000	&OpenFOAM 	&$90^\circ	$&5.913\\
			$1^{*}$	&4.167	&3000	&OpenFOAM 	&$90^\circ	$&4.400\\
			\hline
			2&	41.66&	3000	&OpenFOAM 	&$40^\circ$&	13.56\\
			2&	41.66&	3000	&OpenFOAM 	&$50^\circ$&	12.66\\
			2&	41.66&	3000	&OpenFOAM 	&$50^\circ$&	12.72\\
			2&	41.66&	3000	&OpenFOAM 	&$60^\circ$&	12.66\\
			2&	41.66&	3000	&OpenFOAM 	&$70^\circ$&	12.60\\
			2&	41.66&	3000	&OpenFOAM 	&$80^\circ$&	11.93\\
			2&	41.66&	3000	&OpenFOAM 	&$90^\circ$&	11.20\\
			2&	41.66&	3000	&OpenFOAM 	&$100^\circ$&	9.966\\
			2&	41.66&	3000	&OpenFOAM 	&$110^\circ$&	9.300\\
			2&	41.66&	3000	&OpenFOAM 	&$120^\circ$&	8.987\\
			2&	41.66&	3000	&OpenFOAM 	&$130^\circ$&	8.879\\
			\hline
			3&	41.67&	7500&	OpenFOAM	&$90^\circ$&	12.94\\
			3&	41.67&	5000&	OpenFOAM	&$90^\circ$&	12.63\\
			3&	41.67&	3750&	OpenFOAM	&$90^\circ$&	11.7828	\\
			3&	41.67&	3000&	OpenFOAM	&$90^\circ$&	11.20\\
			3&	41.67&	1500&	OpenFOAM	&$90^\circ$&	9.496\\
			3&	41.67&	750&	OpenFOAM	&$90^\circ$&	8.266\\
			3&	41.67&	500&	OpenFOAM	&$90^\circ$&	7.639\\
			\hline
\end{tabular}
\caption{Summary of the results of the simulation campaigns.  The asterisk indicates a simulation where the predicted value of $\rr$ is negative, corresponding to a low Weber number where the correlation~\eqref{eq:predicted} does not apply.}
\label{tab:camp1}
\end{table}
\begin{table}
	\centering
\begin{tabular}{|p{2cm}|p{1.5cm}|p{1.5cm}|p{2.5cm}|p{1.5cm}|p{2cm}|}
\hline
Campaign & $\myWe$ & $\myRe$ & Method & $\thetaadv$ & $\rr/R_0$ \\
\hline 
\hline		
			$4^{*}$&	1&	577.9&	OpenFOAM	&$90^\circ$&	2.333\\
			4&	10&	577.9&	OpenFOAM	&$90^\circ$&	5.066\\
			4&	41&	577.9&	OpenFOAM	&$90^\circ$&	7.500\\
			4&	100&	577.9&	OpenFOAM	&$90^\circ$&	8.666\\
			4&	1000&	577.9&	OpenFOAM	&$90^\circ$&	8.916\\
			\hline
			$5^{*}$&	0.05&	577.9&	DIM&$	90^\circ$&	2.10\\
			$5^{*}$&  0.1&	577.9&	DIM&$	90^\circ$&	2.15\\
			$5^{*}$&	0.2&	577.9&	DIM&$	90^\circ$&	2.20\\
			$5^{*}$&	1&	  577.9&	DIM&$	90^\circ$&	2.54\\
			5&	10&	  577.9&	DIM&$	90^\circ$&	5.00\\
			5&	20&	  577.9&	DIM&$	90^\circ$&	6.05\\
			5&	100&	577.9&	DIM&$	90^\circ$&	7.65\\
			5&	1000&	577.9&	DIM&$	90^\circ$&	8.55\\
			\hline
		  6&	41.66&	7500	&OpenFOAM 	&$60^\circ$&	15.98\\
			6&	41.66&	7500	&OpenFOAM 	&$70^\circ$&	15.26\\
			6&	41.66&	7500	&OpenFOAM 	&$80^\circ$&	14.97\\
			6&	41.66&	7500	&OpenFOAM 	&$90^\circ$&	12.94\\
			6&	41.66&	7500	&OpenFOAM 	&$100^\circ$&	12.05\\
			6&	41.66&	7500	&OpenFOAM 	&$110^\circ$&	11.22\\
			6&	41.66&	7500	&OpenFOAM 	&$120^\circ$&	10.54\\
			\hline
\end{tabular}
\caption{Summary of the results of the simulation campaigns (continued).  The asterisk indicates a simulation where the predicted value of $\rr$ is negative, corresponding to a low Weber number where the correlation~\eqref{eq:predicted} does not apply.  Campaign 5 uses the Diffuse Interface Method (DIM) for the numerical simulations.}   
\label{tab:camp2}
\end{table}

\section{Discussion and Conclusions}
\label{sec:conc}

We have formulated a rim-lamella model of droplet spreading valid for 2D droplets.  The model may be of use in laboratory setups where cylindrical (quasi-two-dimensional) droplets may be engineered, and also as a theoretical tool for understanding the collision of liquid sheets.  The model reveals the scaling of the maximum spreading radius with Weber number $\myWe$ and Reynolds number $\myRe$.  For the inviscid case, the rim-lamella model reveals a scaling behavior $\rr/R_0\sim \myWe$, a result consistent with energy-budget analyses and numerical simulation~\citep{wu2021decoupled}.  For the viscous case, the results of various simulation campaigns involving systematic variation of $\myWe$, $\myRe$, and advancing contact angle $\thetaadv$ reveal that the data are well approximated by a relationship of the form $\rr/R_0 \sim k_1\myRe^{1/3}-k_0(1-\cos\thetaadv)^{1/2}(\myRe/\myWe)^{1/2}$.  These numerical results are consistent with the theoretical bounds which we have derived for the rim-lamella model using Gronwall's inequality, namely 
$k_1\myRe^{1/3}-k_1(1-\cos\thetaadv)^{1/2}(\myRe/\myWe)^{1/2}\leq \rr/R_0\leq k_1 \myRe^{1/3}$.  Here, $k_0$, $k_1$, and $k_2$ are $O(1)$ constants, which we have carefully estimated.

These results for the 2D droplets are analogous to bounds obtained for 3D axisymmetric droplets~\citep{naraigh2025bounds} as well as to the semi-empirical correlation for droplet spreading obtained in Reference~\cite{roisman2009inertia}.  The semi-empirical correlation can be understood theoretically as containing a positive contribution from the viscous film spreading ($\sim \myRe^{1/5}$ in 3D), as well as a negative contribution which retards the droplet spreading, due to surface tension.  The negative contribution is proportional to $c(t_*)t_*$, where $c(t)$ is a characteristic speed  the functional form of which is similar to the Taylor--Culick speed, and $t_*$ is the time at which the boundary layer reaches the lamella height.  Using a scaling argument for 3D axisymmetric droplets, one obtains $c(t_*)t_*\sim \myRe^{2/5}/\myWe^{1/2}$.  By applying the same arguments in for 2D droplets, one obtains a positive contribution to the droplet spreading proportional to $\myRe^{1/3}$, and a negative contribution proportional to $(\myWe/\myRe)^{1/2}$, consistent with our arguments based on numerical simulations and the rim-lamella model.

Summarizing, while 2D Cartesian droplets (equivalently, 3D cylindrical droplets) are a niche experimental topic, the corresponding mathematical model can be used to understand certain 3D phenomena, such as the impact of retracting liquid sheets~\citep{neel2020fines}.   The present study sheds light on the scaling behaviour of $\rr/R_0$ (dependence on $\myWe$ and $\myRe$) in the 2D geometry.
Further, the 2D model admits analytical solutions in the inviscid case, and is amenable to simple estimates via Gronwall's inequality in the viscous case.  Hence, the  2D case can be regarded as a useful test-bed for understanding the physics of droplet impact.

\subsection*{Acknowledgments}

We acknowledge the Research IT HPC Service at University College Dublin for providing computational facilities and support that contributed to the research results reported in this paper.
NY was supported by the UCD School of Mathematics and Statistics through a funded summer research placement.  LON was supported by Research Ireland (Dublin) and the Department of Agriculture, Food and Marine (Dublin) under grant 21/RC/10303\_P2 (VistaMilk Phase 2).

\appendix

\section{Mathematical Analysis of the Rim-Lamella Model: Inviscid Case}
\label{sec:app:mathanal_invisc}

We write down the rim-lamella model in the inviscid case:
\begin{subequations}
\begin{eqnarray}
\frac{\mathd V}{\mathd t}&=&2\left(u_0-U\right)h,\\
\frac{\mathd R}{\mathd t}&=&U,\\
V\frac{\mathd U}{\mathd t}&=&2\left(u_0-U\right)^2h-2\gamma\left(1-\cos\thetaadv\right),
\end{eqnarray}
\end{subequations}
where $u_0=R/(t+t_0)$.  We let $\Delta=u_0-U$, hence
\begin{equation}
\frac{\mathd \Delta}{\mathd t}+\frac{\Delta}{t+t_0}=-\frac{2h}{V}\left(\Delta^2-c^2\right),
\end{equation}
where $c^2=[\gamma/(\rho h)](1-\cos\thetaadv)$.  We look at $V(\mathd \Delta/\mathd t)+\Delta(\mathd V/\mathd t)$.  This reduces down to:
\begin{equation}
\frac{\mathd }{\mathd t}(V\Delta)+\frac{V\Delta}{t+t_0}=2hc^2=2(\gamma/\rho)(1-\cos\thetaadv).
\end{equation}
Thus, there is an exact solution for $V\Delta$:
\begin{equation}
V\Delta=V_{init}\Delta_{init}\left(\frac{\tau+t_0}{t+t_0}\right)+\frac{\gamma}{\rho}\left(1-\cos\thetaadv\right)\left[t+t_0-\frac{(\tau+t_0)^2}{t+t_0}\right].
\end{equation}
We recall the definition of $\Delta$, an re-write this as:
\begin{equation}
\frac{\mathd R}{\mathd t}-\frac{R}{t+t_0}=-\frac{V_{init}}{V}\Delta_{init}\left(\frac{\tau+t_0}{t+t_0}\right)-\frac{\gamma}{\rho}\frac{\left(1-\cos\thetaadv\right)}{V}\left[t+t_0-\frac{(\tau+t_0)^2}{t+t_0}\right].
\end{equation}
This can be further re-written as:
\begin{equation}
\frac{\mathd }{\mathd t}\left(\frac{R}{t+t_0}\right)=-\frac{V_{init}}{V}\frac{\Delta_{init}}{t+t_0}\left(\frac{\tau+t_0}{t+t_0}\right)-\frac{\gamma}{\rho}\frac{\left(1-\cos\thetaadv\right)}{V}\left[1-\left(\frac{\tau+t_0}{t+t_0}\right)^2\right].
\end{equation}
As in the main part of the paper, we use the remote asymptotic solution for $h$, $h(R,t)=[(\tau+t_0)/(t+t_0)]h_{init}$, hence $V=V_{tot}-2R[(\tau+t_0)/(t+t_0)]h_{init}$.  We will group together the terms $R[(\tau+t_0)/(t+t_0)]$ as $\eta$.  Thus, we have:
\begin{equation}
\frac{\mathd\eta}{\mathd t}=-\Delta_{init}\frac{V_{init}}{V_{tot}-2\eta h_{init}}\left(\frac{\tau+t_0}{t+t_0}\right)^2-\frac{\gamma(\tau+t_0)}{\rho}\frac{\left(1-\cos\thetaadv\right)}{V_{tot}-2\eta h_{init}}\left[1-\left(\frac{\tau+t_0}{t+t_0}\right)^2\right].
\end{equation}
This is a separable ODE:
\begin{equation}
\left[V_{tot}-2\eta h_{init}\right]\mathd \eta=\bigg\{-\Delta_{init}V_{init}\left(\frac{\tau+t_0}{t+t_0}\right)^2-\frac{\gamma(\tau+t_0)}{\rho}\left(1-\cos\thetaadv\right)\left[1-\left(\frac{\tau+t_0}{t+t_0}\right)^2\right]\bigg\}.
\end{equation}
Therefore:
\begin{multline}
V_{tot}\eta-h_{init}\eta^2=V_{tot}\eta_{init}-h_{init}\eta_{init}^2\\-
\underbrace{\bigg\{\frac{\gamma(\tau+t_0)}{\rho}\left(1-\cos\thetaadv\right)\left[t+\frac{(\tau+t_0)^2}{t+t_0}\right]-\Delta_{init}V_{init}(\tau+t_0)\left(\frac{\tau+t_0}{t+t_0}\right)\bigg\}}_{=G(t)}+G(\tau),
\end{multline}
where $\eta_{init}=R_{init}$.
In simple terms, we now have:
\begin{equation}
h_{init}\eta^2-V_{tot}\eta=h_{init}\eta_{init}^2-V_{tot}\eta_{init}+[G(t)-G(\tau)],
\end{equation}
or
\begin{equation}
h_{init}\eta^2-V_{tot}\eta-h_{init}\eta_{init}^2+V_{tot}\eta_{init}-[G(t)-G(\tau)]=0.
\end{equation}
Hence, since $\eta=R[(\tau+t_0)/(t+t_0)]$, we have:
\begin{equation}
R\left(\frac{\tau+t_0}{t+t_0}\right)=\frac{V_{tot}\pm \sqrt{V_{tot}^2-4h_{init}\bigg\{ -h_{init}\eta_{init}^2+V_{tot}\eta_{init}-[G(t)-G(\tau)]\bigg\}}}{2h_{init}}.
\label{eq:app:Rx1}
\end{equation}
Since $V_{tot}=V_{init}+2R_{init}h_{init}$, we have:
\begin{eqnarray*}
V_{tot}^2+4h_{init}^2\eta_{init}^2-4h_{init}V_{tot}\eta_{init}&=&\left(V_{tot}-2h_{init}\eta_{init}\right)^2,\\
                                                    &=&\left(V_{tot}-2h_{init}R_{init}\right)^2,\\
																										&=&V_{init}^2.
\end{eqnarray*}
Hence, Equation~\eqref{eq:app:Rx1} becomes:
\begin{equation}
R\left(\frac{\tau+t_0}{t+t_0}\right)=\frac{V_{tot}\pm \sqrt{ V_{init}^2+4h_{init}[G(t)-G(\tau)]\bigg\}}}{2h_{init}}
\end{equation}
This can be re-written further:
\begin{equation}
\underbrace{\frac{R}{V_{tot}/(2h_{init})}}_{=Y}=X\pm X\sqrt{ \epsilon^2+4(h_{init}/V_{tot})[G(t)-G(\tau)]},\qquad \epsilon=\frac{V_{init}}{V_{tot}},\qquad X=\frac{t+t_0}{\tau+t_0}.
\end{equation}
This simplifies again:
\begin{multline}
Y=X\\
\pm X\sqrt{\epsilon^2+\frac{4h_{init}}{V_{tot}^2}\frac{\gamma(\tau+t_0)^2}{\rho}\left(1-\cos\thetaadv\right)\left(X+\frac{1}{X}-2\right)+\frac{4\Delta_{init}(\tau+t_0)V_{init}h_{init}}{V_{tot}^2}\left(1-\frac{1}{X}\right)},
\end{multline}
valid for $X\geq 1$.

We identify a characteristic speed $c_*^2=4(h_{init}/V_{tot})(\gamma/\rho)(1-\cos\thetaadv)$ and hence a characteristic lengthscale $\ell_*^2=c_*^2(\tau+t_0)^2$ and hence $A=\ell_*^2/V_{tot}$ is dimensionless.  Similarly, $4\Delta_{init}(\tau+t_0)h_{init}/V_{tot}=B$ is dimensionless, so we are left with:
\begin{equation}
Y=X\pm X\sqrt{\epsilon^2+A \left(X+\frac{1}{X}-2\right)+ \epsilon B \left(1-\frac{1}{X}\right)}.
\end{equation}
Thus, we are left with an expression involving only polynomials and square roots of polynomials:
\begin{equation}
Y=X\pm \sqrt{\epsilon^2 X^2+A \left(X^3+X-2X^2\right)+\epsilon B\left(X^2-X\right)}.
\label{eq:app:YX}
\end{equation}
For $Y(X)$ (hence $R(t)$) to have a maximum value, the negative sign in front of the square root must be chosen.  Thus, Equation~\eqref{eq:exact1} in the main paper is obtained.

\subsection*{Maximum Spreading -- Inviscid Case}

The maximum spreading occurs when $\mathd R/\mathd t=0$, hence $\mathd Y/\mathd X=0$.  Referring to Equation~\eqref{eq:app:YX}, this occurs when
\begin{equation}
4\left[\epsilon^2 X^2+A \left(X^3+X-2x^2\right)+B \epsilon\left(X^2-X\right)\right]=
\left[3A X^2+2x\left(\epsilon^2+\epsilon B-2A\right)+\left(A-\epsilon B\right)\right]^2.
\end{equation}
An asymptotic solution valid for large $X$ can be obtained by balancing the highest powers of $X$ on each side of this polynomial equation:
\begin{equation}
4A X^3\approx 9 A^2 X^4,
\end{equation}
Hence, $X\sim 4/(9A)$, for $X\gg 1$.  Plugging in the numbers gives
\begin{equation}
Y_{max}\sim (4/9)A^{-1}-\sqrt{A (4/9)^3 A^{-3}}=\tfrac{4}{27}A^{-1},\qquad X\gg 1.
\label{eq:app:Ymax}
\end{equation}
By using the explicit definition of $A$ in terms of $\gamma$, etc. we have, in the large-$\myWe$ limit:
\begin{eqnarray}
\frac{R_{max}}{R_0}&=& Y_{max}\left(\frac{V_{tot}}{2h_{init}}\right),\nonumber\\
									 &\stackrel{\text{Eq.~\eqref{eq:app:Ymax}}}{\sim }&  \tfrac{4}{27}\frac{V_{tot}^2}{4 h_{init} (\gamma/\rho)(1-\cos\thetaadv)(\tau+t_0)^2}\left(\frac{V_{tot}}{2h_{init}}\right),\qquad X\gg 1.
\end{eqnarray}
This establishes Equation~\eqref{eq:rmaxWe} in the main part of the paper.
\section{Mathematical Analysis of the Rim-Lamella Model: Viscous Case}
\label{sec:app:mathanal_visc}

In this Appendix we present a detailed analysis of the rim-lamella model in the viscous case, the main points of which are presented in  Section~\ref{sec:viscous}.  The starting-point is the basic rim-lamella modle in the viscous case (Equations~\eqref{eq:RLvisc1}).  We introduce the velocity defect
\begin{equation}
\Delta=u_0\left(1-\frac{\hbl}{h}\right)-U;
\end{equation}
we thereby reduce Equations~\eqref{eq:RLvisc1} to a simpler form:
\begin{subequations}
\begin{eqnarray}
\frac{\mathd V}{\mathd t}&=&2\Delta h,\\
V\frac{\mathd U}{\mathd t}&=&2(\Delta^2-c^2)h.
\end{eqnarray}%
\label{eq:app:RLvisc2}%
\end{subequations}%
By direct computation, we get:
\begin{equation}
\frac{\mathd\Delta}{\mathd t}+\frac{\Delta}{t+t_0}\left(1-\frac{\hbl}{h}\right)=-2(\Delta^2-c^2)(h/V)-\frac{u_0}{t+t_0}\frac{\hbl}{h}\underbrace{\left[2\left(1-\frac{\hbl}{h}\right)+\tfrac{1}{2}\frac{t+t_0}{t+t_1}\right]}_{=\Phi(t)\geq 0}.
\end{equation}
We therefore obtain:
\begin{equation}
\frac{\mathd }{\mathd t}(\Delta V)+\frac{\Delta V}{t+t_0}\left(1-\frac{\hbl}{h}\right)=2(\gamma/\rho)(1-\cos\thetaadv)-\frac{u_0}{t+t_0}(\hbl/h)\Phi(t).
\end{equation}
We identify the integrating factor 
\begin{equation}
\mu=\mathe^{\int \frac{1}{t+t_0}[1-(\hbl/h)]\mathd t}=\mathe^{-\int (1/h)(\mathd h/\mathd t)\mathd t}=\mathe^{- \int\mathd h/h}=1/h.
\end{equation}
Hence:
\begin{equation}
\frac{\mathd }{\mathd t}\left(\frac{\Delta V}{h}\right)=2(\gamma/\rho)(1-\cos\thetaadv)(1/h)
-\frac{u_0}{t+t_0}(\hbl/h^2)\Phi(t)\leq 2(\gamma/\rho)(1-\cos\thetaadv)(1/h).
\end{equation}
We call
\begin{equation}
I_h(t)=\int_\tau^t \frac{\mathd t}{h}.
\end{equation}
With $V(t=\tau)=0$, we have:
\begin{equation}
\frac{\Delta V}{h} \leq 2(\gamma/\rho)(1-\cos\thetaadv)I_h(t).
\end{equation}
Or:
\begin{equation}
\Delta \leq (\gamma/\rho)(1-\cos\thetaadv)(h/V)I_h(t).
\end{equation}
But $\Delta=[R/(t+t_0)][1-(\hbl/h)]-(\mathd R/\mathd t)$, hence:
\begin{equation}
\frac{\mathd R}{\mathd t}-\frac{R}{t+t_0}\left(1-\frac{\hbl}{h}\right)\geq -2(\gamma/\rho)(1-\cos\thetaadv)(h/V)I_h(t).
\end{equation}
The integrating factor here is $h$, so by Gronwall's Inequality, we have:
\begin{equation}
\frac{\mathd }{\mathd t}(Rh)\geq -2(\gamma/\rho)(1-\cos\thetaadv)(h^2/V)I_h(t).
\end{equation}
We have $V=V_{tot}-2Rh$.  So we identify $\eta=Rh$ to get:
\begin{equation}
\frac{\mathd \eta}{\mathd t}\geq -2(\gamma/\rho)(1-\cos\thetaadv)\frac{h^2 I_h(t)}{V_{tot}-2\eta}.
\end{equation}
Since $V_{tot}-2\eta =V\geq 0$, we can multiply up without violating the inequality, to get:
\begin{equation}
\frac{\mathd}{\mathd t}\left(V_{tot}\eta-\eta^2\right)\geq 2(\gamma/\rho)(1-\cos\thetaadv) h^2 I_h(t).
\end{equation}
Hence:
\begin{equation}
V_{tot}\eta-\eta^2\geq V_{tot}\eta_{init}-\eta_{init}^2 -(\gamma/\rho)(1-\cos\thetaadv)
\underbrace{\left[2\int_\tau^t h^2 I_h(t)\mathd t\right]}_{=\Delta G(t)}.
\end{equation}
This is a quadratic inequality.  Critical points occur at $\eta_{\mathrm{cr},\pm}(t)$, where:
\begin{equation}
\eta_{\mathrm{cr},\pm}(t)=\frac{V_{tot}\pm \sqrt{ V_{tot}^2-4\left[-(\gamma/\rho)(1-\cos\thetaadv)[\Delta G(t)]-\eta_i^2+V_{tot}\eta_i\right]}}{2}.
\end{equation}
The square root can be simplified:
\begin{equation}
\eta_{\mathrm{cr},\pm}(t)=\frac{V_{tot}\pm \sqrt{ V_i^2+4(\gamma/\rho)(1-\cos\thetaadv)[\Delta G(t)]}}{2}.
\end{equation}
When $V_i=0$ this simplifies again:
\begin{equation}
\eta_{\mathrm{cr},\pm}(t)=\tfrac{1}{2}V_{tot}\pm \left[(\gamma/\rho)(1-\cos\thetaadv)\right]^{1/2}[\Delta G(t)]^{1/2}
\end{equation}
Thus, $R(t)h(t)\leq \eta_{\mathrm{cr},+}(t)$ with the plus sign chosen, hence:
\begin{equation}
R(t)\leq \frac{\tfrac{1}{2}V_{tot}+\left[(\gamma/\rho)(1-\cos\thetaadv)\right]^{1/2}[\Delta G(t)]^{1/2}}{h(t)}.
\end{equation}
This is an analytical upper bound for $R(t)$.  Similarly, with the negative sign chosen, we get:
\begin{equation}
R(t)\geq \frac{\tfrac{1}{2}V_{tot}-\left[(\gamma/\rho)(1-\cos\thetaadv)\right]^{1/2}[\Delta G(t)]^{1/2}}{h(t)}.
\end{equation}
We have:
\begin{multline}
\max R(t)\geq \max \left[\frac{\tfrac{1}{2}V_{tot}-\left[(\gamma/\rho)(1-\cos\thetaadv)\right]^{1/2}[\Delta G(t)]^{1/2}}{h(t)}\right]\\
\geq \frac{\tfrac{1}{2}V_{tot}-\left[(\gamma/\rho)(1-\cos\thetaadv)\right]^{1/2}[\Delta G(t_*)]^{1/2}}{h_*}.
\end{multline}
For the upper bound, we reuse the argument in Reference~\cite{naraigh2025bounds}.  We have:
\begin{equation}
2\underbrace{R_{max}}_{=R(t_{max})}h(t_{max})=V_{tot}-V(t_{max})\leq V_{tot}.
\end{equation}
Hence,
\begin{equation}
R_{max}\leq \frac{V_{tot}}{2 h(t_{max})}.
\end{equation}
If the maximum occurs in Phase 1, then $h(t)$ is monotone-decreasing, hence $h(t_{max})\geq h_*$, hence
$1/h(t_{max})\leq 1/h_*$, hence:
\begin{equation}
R_{max}\leq \frac{V_{tot}}{2 h_*}.
\end{equation}
If the  maximum occurs in Phase 2, then $h(t)$ is a  constant function and equal to $h_*$, so it is still the case that $R_{max}\leq V_{tot}/(2h_*)$.  Hence, in both cases, we have:
\begin{equation}
R_{max}\geq R_* \geq V_{tot}/(2h_*).
\end{equation}
 Hence, by a sandwich result, we have:
\begin{equation}
\frac{\tfrac{1}{2}V_{tot} - \left[(\gamma/\rho)(1-\cos\thetaadv)\right]^{1/2}[\Delta G(t_*)]^{1/2}}{h_*} \leq \max_t R(t)\leq \frac{V_{tot}}{2 h_*}
\label{eq:app:Rineq3}
\end{equation}
This establishes Equation~\eqref{eq:Rineq3} in the main paper.



\end{document}